\documentclass[10pt-default,11pt]{article}
\usepackage{amsmath}
\usepackage{amsfonts}
\usepackage{amssymb}
\usepackage{geometry}

\input{tcilatex}
\begin{document}

{\Large How the Modified Bertrand Theorem Explains Regularities}

{\Large and Anomalies} {\Large of the Periodic Table of Elements\bigskip }

\textbf{Arkady L. Kholodenko\footnote{{\large Correspondence: }%
string@clemson.edu}}$\bigskip $

Department \ of Chemistry, Clemson \ University, Clemson, SC 29634-0973,
United States

\ \ \ \ \ \ \ \ \ \ \ \ \ \ \ \ \ \ \ \ \ \ \ \ \ \ \ \ \ \ \ \ \ \ \ \ \ \
\ \ \ \ \ \ \ \ \ \ \ \ \ \ \ \ \ \ \ \ \ \ \ \ \ \ \ \ \ \ \ \ \ \ \ \ \ \
\ \ \ \ \ \ \ \ \ \ \ \ \ \ \ \ \ \ \ \ \ \ \ \ \ \ \ \ \ \ \ \ \ \ \ \ \ \
\ \ \ \ \ \ \ \ \ \ \ \ \ \ \ \ \ \ \ \ \ \ \ \ \ \ \ \ \ \ \ \ 

\bigskip

\ \ \ \ \ \ \ \ \ \ \ \ \ \ \ \ \ \ \ \ \ \ \ \ \ \ \ \ \ \ \ \ \ \ \ \ \ \
\ \ \ \ \ \ \ \ \ 

Bertrand theorem permits closed orbits in 3d Euclidean space only for 2
types of central

potentials. These are of Kepler- Coulomb and harmonic oscillator type.
Volker Perlick \ 

recently extended Bertrand theorem. He designed new static spherically
symmetric

(Bertrand) spacetimes obeying Einstein's equations and supporting closed
orbits. In this

work we demonstrate that the topology and geometry of these spacetimes
permits us to

solve quantum many-body problem for any atom of periodic system exactly. The

computations of spectrum for any atom are analogous to that for hydrogen
atom. Initially,

the exact solution of the Schr\"{o}dinger equation for any

multielectron atom (without reference to Bertrand theorem) was obtained by
Tietz in

1956. We recalculated Tietz results by applying the methodology consistent
with

new (different from that developed by Fock in 1936) way of solving Schr\"{o}%
dinger's

equation for hydrogen atom. By using this new methodology it had become
possible

to demonstrate that the Tietz- type Schr\"{o}dinger's equation is in fact
describing

the quantum motion in Bertrand spacetimes. As a bonus, we solved analytically

the L\"{o}wdin's challenge problem. Obtained solution is not universal
though since

there are exceptions of the Madelung rule in transition metals and among

lanthanides and actinides. Quantum mechanically these exceptions as well as

the rule itself are treated thus far with help of relativistic Hartree-Fock
calculations.

The obtained results do not describe the exceptions in detail yet. However,

studies outlined in this paper indicate that developed new methods are
capable of

describing exceptions as well. The paper ends with some remarks about
usefulness

of problems of atomic physics for development of quantum mechanics, quantum

field theory and (teleparallel) gravity

\bigskip

\bigskip \textrm{Keywords:\ \bigskip\ Maxwell's fish-eye potential}$,$ 
\textrm{conformally invariant classical and quantum}

\bigskip

\textrm{equations}$,$ \textrm{quantum Bertrand spacetimes\bigskip \bigskip
\medskip , Madelung rule, Einstein-Cartan and teleparallel}

\textrm{gravity}

\ \ \ \ \ \ \ \ \ \ \ \ \ \ \ \ \ \ \ \ \ \ \ \ \ \ \ \ \ \ \ \ \ \ \ \ \ \
\ \ \ \ \ \ \ \ \ \ \ \ \ \ \ \ \ 

{\Large Background and summary }\textbf{\bigskip\ \ \ \ }

\textbf{Overview of trends in the description of periodic system of elements.%
}

\textbf{Statement of \ L\"{o}wdin's challenge problem}\medskip

Although quantum mechanical description of multielectron atoms and molecules
is

considered to be a well established domain of research, recently published
book

(Thussen and Ceulemans, 2017) indicates that there are still many things
which,

fortunately, are left for further development. Even though the cited book
represents

a significant step toward improvement of the existing description of
electronic structure

of atoms and molecules, we were able to find many items requiring further
study.

Specifically, the quantum mechanical description of multielectron atom (with

atomic number $Z$ and infinitely heavy nucleus) begins with writing down the

stationary Schr\"{o}dinger equation 
\begin{equation}
\hat{H}\Psi (\mathbf{r}_{1},\mathbf{r}_{2},...,\mathbf{r}_{Z})=E\Psi (%
\mathbf{r}_{1},\mathbf{r}_{2},...,\mathbf{r}_{Z})  \tag{1}
\end{equation}

with the Hamiltonian%
\begin{equation}
\hat{H}=-\dsum\limits_{i=1}^{Z}\frac{\hslash ^{2}}{2m}\nabla
_{i}^{2}-\dsum\limits_{i=1}^{Z}\frac{Ze^{2}}{r_{i}}+\frac{1}{2}\dsum\limits 
_{\substack{ i,j=1  \\ i\neq j}}^{Z}\frac{e^{2}}{r_{ij}}.  \tag{2}
\end{equation}

Following Bohr's \textsl{Aufbauprinzip} the atom with atomic number Z is
made up of electrons

added in succession to the bare atomic nucleus. At the initial stages of
this process

electrons are assumed to occupy the one-electron levels of lowest energy.
This process is

described in terms of the one electron eigenvalue problem 
\begin{equation}
\hat{H}_{i}\psi _{\square _{i}}(\mathbf{r}_{i})=[-\frac{\hslash ^{2}}{2m}%
\nabla _{i}^{2}+V_{eff}(\mathbf{r}_{i})]\psi _{\square _{i}}(\mathbf{r}%
_{i})=\varepsilon _{nl}(i)\psi _{\square _{i}}(\mathbf{r}_{i}),i=1\div Z, 
\tag{3}
\end{equation}

where $V_{eff}(\mathbf{r}_{i})$ is made of the combined nuclear potential - $%
\frac{Ze^{2}}{r_{i}}$ for the i-th electron and

the centrally symmetric Hartree-Fock type potential $\mathcal{F}$(\textbf{r}$%
_{i})$ \ coming from the presence

of the rest of atomic electrons. The fact that $\mathcal{F}$(\textbf{r}$%
_{i}) $ is indeed centrally symmetric \ was

discussed in the book by Bethe and Jackiw (Bethe and Jackiw\textsl{, }2018).
It is fundamentally

important for our calculations. The symbol $\square _{i}$ indicates the i-th
entry into the set\ made

of hydrogen-like quantum numbers characterizing individual electrons. Recall
that the

concept of\textsl{\ orbital} is determined by the major quantum number $n$
having its origin in

studies of hydrogen atom. The number of electrons allowed to sit on a given
orbital is

determined by the \textsl{Pauli exclusion principle. }Thus, with increasing
Z electrons are

occupying successive orbitals according to Bohr's Aufbau scheme until the 
\textsl{final }

\textsl{ground state electron configuration} is reached. \ Since electrons
are indistinguishable,

the hydrogen-like quantum numbers $n,l,m$ and $m_{s}$ cannot be associated
with a particular

electron. Therefore, the symbol $\ \square _{i\text{ }}$ should be
understood as representing a specific set

of quantum numbers otherwise used for description of individual (that is not
collectivized)

electrons.The problem with just described \textsl{Aufbauprinzip }lies in the
assumption that

the guiding principle in designing the \ final ground state electron
configuration is made

out of two components: a) knowledge of hydrogen atom-like wave functions
supplying

the quantum boxes/numbers $\square _{i}$ and, b) the Pauli principle which
is mathematically

restated in the form of the fully antisymmetric wavefunction $\Psi (\mathbf{r%
}_{1},\mathbf{r}_{2},...,\mathbf{r}_{Z})$. Should

these requirements be sufficient, then it would be possible with a good
accuracy

to replace $V_{eff}(\mathbf{r}_{i})$ by -$\frac{Ze^{2}}{r_{i}}$ \ so that
the filling of electronic levels would occur according

to the Fock $n$-rule\medskip

\textbf{Fock n-rule: }\textsl{With increasing Z the nl orbitals are filled
in order of increasing n.\medskip }

This rule leads to the problems already for the lithium as explained in the
book by

Thussen and Ceulemans (Thussen and Ceulemans, 2017), page 330. As result, the

n-rule was replaced by the ($n,l$) rule.\medskip

\textbf{The hydrogenic (}\textit{n},\textit{l})\textbf{\ rule: }\textsl{With
increasing Z, the orbitals are filled in order of }

\textsl{increasing n while for a fixed n the orbitals are filled in order of
increasing \textit{l}.\medskip }

After $Z=18$ the ($n,l$) rule breaks down though. Therefore, it was
subsequently replaced

by the ($n+l,n$) rule suggested by Madelung- the person who reformulated

Schr\"{o}dinger's equation in hydrodynamic form (Kholodenko and Kauffman,
2018).\medskip

\textbf{The Madelung (}\textit{n+l},\textit{n}\textbf{) rule: }\textsl{With
increasing Z, the orbitals are filled in order of }

\textsl{increasing n+}$\QTR{sl}{l}=N.$ \textsl{For fixed }$N$\textsl{, the
orbitals are filled in order of increasing n.\medskip }

All the above rules are empirical. As such, they require theoretical
explanation.

This fact brings us to the\medskip

\textbf{L\"{o}wdin's challenge problem}: Find a way to derive the Madelung
rule \ ab initio. \textsl{\footnote{%
This problem was posed by Per-Olov L\"{o}wdin (L\"{o}wdin,1969). Additional
details and references are in (Allen and Knight, 2002).}\medskip }

The essence of Mendeleev's periodic system of elements lies exactly in
discovered

periodicity of properties of chemical elements. Although there are 100's of
ways

this periodicity can be exhibited\footnote{%
Results of www searches indicate that this process is still ongoing.}, the
commonly accepted periodic table

of elements consists of \ seven periods: 2-8-8-18-18-32-32. Notice that all
period

lengths occur in pairs (period doubling), except for the very first period
of size 2.

To determine whether this exception is intrinsic or not, the analysis of
work by

Charles Janet on \ periodic table done in 1930 (6 years before work by
Madelung!)

is the most helpful. It is summarized in the book by Thussen and Ceulemans,

pages 336-340. Although initially Janet developed his version of periodic
table

without guidance of quantum mechanics, eventually he did make a connection
with

Bohr's results. Janet's periodic table has 8 periods. The periods in Janet's
table

are characterized (without exception) by the constant value of $N=n+l$ in
perfect

agreement \ with the Madelung rule. This fact suggests elevation of the
number

$N=n+l$ to the rank of new quantum number. By organizing the elements in
periods

of constant $n+l$ and groups of constant $l,m_{l}$ and $m_{s}$, the period
doubling emerges

naturally and leads to the sequence of periods:
2-2-8-8-18-18-32-32.\bigskip\ Using apparatus

of the dynamical group theory\ \medskip\ Thussen and Ceulemans were able to
reobtain Janet

sequence. Application of \ group-theoretic analysis to the periodic system
of elements

was done repeatedly in the past. Many references to these earlier works can
be found

in (Thussen and Ceulemans, 2017). To our knowledge, the most notable are
results

presented in Chapter 6 of the book by Englefield (Englefield,1972). The
results of

Chapter 6 are independently reobtained in (Thussen and Ceulemans, 2017).
Should the

Madelung rule be without exceptions, just mentioned results would be
sufficient for

solving of the L\"{o}wdin challenge problem. However, the existing
exceptions for

transition metals, lanthanides and actinides indicate that use of the
dynamical group

theory methods alone is not sufficient. As result, in this work we describe
alternative

methods enabling us to explain the Madelung rule and its exceptions using
different

physical arguments. This had become possible by elaborating on works by
Demkov

and Ostrovsky summarized below.\medskip

\textbf{\ Works by Demkov and Ostrovsky\medskip \medskip\ }

A concise and convincing explanation of the period doubling and its
connection with

the Madelung rule is given in (Scerri and Restrepo, 2018). The origins of
the Madelung

rule had attracted attention of Demkov and Ostrovsky (Demkov and Ostrovsky,
1971).

The impact of their work is of major importance for us. Nevertheless,
subsequently

group-theoretic studies by Kitagawara and Barut (Kitagawara and Barut, 1983)
and,

later in (Kitagawara and Barut,1984) uncovered the apparent flaws in the
logic of

Demkov-Ostrovsky calculations. In their book Thussen and Ceulemans

(Thussen and Ceulemans, 2017) also expressed their objections to results of
the

Demkov-Ostrovsky cycle of works. On page 381 of (Thussen and Ceulemans, 2017)

we found the following statement:

"Demkov and Ostrovsky developed an atomic physics model that incorporates

the Madelung rule, but by replacing the quantization of level energies with

quantization of coupling constants at zero energy."\medskip\ 

Furthermore, in (Kholodenko and Kauffman, 2019) we noticed that Demkov and

Ostrovsky, while being able to obtain correct results, had been unable to
provide

their rigorous justification because their effective potential $V_{eff}(%
\mathbf{r}_{i})$ was guessed.

The authors of the book (Thussen and Ceulemans, 2017) concluded as well that,

even though the Demkov -Ostrovsky results do reproduce the Madelung rule

correctly, the way these results were obtained \ cannot be considered as
solution

of the L\"{o}wdin challenge. This circumstance brings us to the
following.\medskip\ \bigskip

{\Large Summary of solved problems\medskip\ }

\textbf{General background\bigskip }

In (Kholodenko and Kauffman, 2019) we demonstrated that the objections
raised in

(Thussen and Ceulemans, 2017), (Kitagawara and Barut, 1983) and (Kitagawara
and

Barut, 1984) are coming from the lack of knowledge of needed mathematical
apparatus

by the physics \ and chemistry community, including works by Demkov and
Ostrovsly.

In mathematical physics community this apparatus is already known in many
other

contexts. Thus, one of the tasks of this work is to introduce this apparatus
to the atomic

and molecular physics community. By doing so a number of \ problems of major

importance can be solved \ so that all of the objections raised in just
cited references

are removed. \ Ourselves, we also found some additional objections. They are
to be

described below and are also removed. This fortunate circumstance is paving
a reliable

way for subsequent study of exceptions. Detailed results are to be presented
in later

works.

In their seminal works Demkov and Ostrovsky (D-O) realized that the key to
success

of solving L\"{o}wdin's problem lies is Eq.(2), where $V_{eff}(\mathbf{r}%
_{i})$ should be chosen correctly.

The Bertrand theorem of classical mechanics (Goldstein, Poole and Safko,
2014)

imposes seemingly insurmountable restrictions on selection of $V_{eff}(%
\mathbf{r}_{i})$ since for

spherically symmetric potentials only the Coulombic -$\frac{Ze^{2}}{r_{i}}$
and the harmonic oscillator

$kr^{2}$ potentials allow dynamically closed orbits. D-O believed that, in
spite of the

indistinguishability of electrons, Bohr's (circular orbits) and, later on,
Sommerfeld's

(elliptical orbits) treatment of hydrogen atom (Sommerfeld, 1934), is
essential for

obtaining the discrete spectrum of multielectron atom since the
semiclassical-classical

methods of treatment of the spectral problem should be associated with
closed orbits. \ 

At the semiclassical level\ of description of multielectron atoms the role
of closed

orbits very recently was discussed, for example in (Akila et al, 2017).
Beginning with

the classical treatment of motion of electrons in helium, the classical
(and, hence,

the semiclassical!) dynamics of \ electrons in multielectron atoms is
believed to be

chaotic. The seminal book by Gutzwiller (Gutzviller, 1990) is an excellent
introduction

to this topic. \bigskip In this work, by developing D-O ideas we argue, that
the Madelung rule

is not a reflection of the chaotic dynamics of electrons in multielectron
atoms. Instead,

it is a reflection of some deep differential geometric and topological
properties intrinsic

for dynamics of electrons in multielectron atoms.\medskip\ Only the
fundamentals are provided

below. The description of mathematical details initiated in

(Kholodenko and Kauffman, 2019) is to be extended in future
publications.\medskip

\textbf{Arguments leading to extension of the classical Bertrand
theorem\bigskip }

Hoping to bypass the limitations of the Bertrand theorem, D-O employed the
optical

-mechanical analogy in their calculations. It permitted them to use the
Maxwell fish-eye

potential (and its conformally deformed modifications), e.g. see Eq.(5)
below.

The Maxwell fish-eye potential, is used instead of the Coulombic potential
for the hydrogen

atom. Its conformally deformed modification is used instead of $V_{eff}(%
\mathbf{r}_{i})$ for multielectron

atoms. At the level of classical mechanics D-O demonstrated (Demkov and

Ostrovsky, 1971) the equivalence (for the hydrogen atom) between the
Hamilton-Jacobi

equations employing \ the Maxwell fish -eye and Coulombic potentials. By
using the

Maxwell fish-eye potential instead of Coulombic, D-O hoped to bypass
limitations of the

Bertrand theorem. They assumed that the conformally modified fish-eye
potential can be

used instead of $V_{eff}(\mathbf{r}_{i})$ for multielectron atoms. Their
work attracted attention of John

Wheeler who in (Wheeler,1971), (Wheeler 1976) and, with his student (Powers,
1971),

studied classically and semiclassically the electron motion in the fish-eye
and conformally

deformed fish-eye potentials. The dynamics of electrons in such conformally
deformed

potentials according to these studies happens to involve orbits \ which are
closed, planar

and have self-intersections$.$In another of his paper (Ostrovsky,1981)
Ostrovsky argued

that the self-intersections of orbits do not contradict the Bertrand
theorem. This

statement by Ostrovsky happens to be wrong. Details are explained in
(Kholodenko and

Kauffman, 2019). Here we provide only very basic arguments.

Classical mechanics treatment of the confined motion of the electron in the
Coulombic

potential indicates that the motion is planar. Use of the stereographic
projection converts

the plane into hemisphere as is well known. There could be conversion to the
northern or

southern hemispheres. Therefore, the plane \textbf{R}$^{2}$ with one extra
point added (the point at

infinity) allows us to identify the plane \textbf{R}$^{2}$(with added extra
point) and the sphere, $S^{2}.$

Now, we consider a "trivial" problem: how to describe all closed curves on
the sphere?

This problem happens to be not as trivial as it looks. It was solved by
Little (Little,

1970). He found that there are only 3 distinct regular homotopy classes of
oriented

closed curves on $S^{2}.$These are: a) those for curves without
self-intersections, b) those

for curves with just 1 self--intersection and, c) those with 2
self-intersections. The

b)-type homotopy class curves were \ obtained by Wheeler (Wheeler,1971). Much

more recently, in 2017, the same self-intersecting patterns were obtained
for dynamical

trajectories existing in Bertrand spacetimes, e.g. see page 3362 of (Kuru et
al, 2017).

Their results were obtained without any reference to atomic physics or
results by

Wheeler. The Kepler-Coulomb dynamics in flat 3d Euclidean space does not
allow

self-intersections. The self-intersections are allowed if the flat space
Bertrand

theorem is extended to the motions on curved (Bertrand) manifolds (Kuru et
al, 2017).\medskip

\textsl{Thus, it follows, that the effects of curvature and the presence of}

\textsl{self-intersections in dynamical trajectories are connected to each
other.\medskip }

The sphere $S^{2}$ is conformally flat. That is to say, there is a
transformation

(a conformal transformation, in fact)\footnote{%
It is known that the stereographic projection is such a transformation.
Whether or not there are other conformal transformations is of no concern to
us in this work.} connecting flat manifolds, e.g. plane \textbf{R}$^{2}$,

with curved manifolds, e.g. sphere, $S^{2}.$\textbf{The results of the
modified Bertrand }

\textbf{theorem are valid exactly for the conformally flat manifolds}. This

fact was proven in (Ballesteros, 2009). Thus, flatness of self-intersecting
patterns obtained

in (Kuru et al, 2017) is, in fact, conformal flatness.

To connect Bertrand spacetimes with atomic physics we begin with D-O
statement made

in (Demkov and Ostrovsky,1971). "The Maxwell's fish-eye problem is \textbf{%
closely related}

\textbf{to} the Coulomb problem." \ Being aware of the book by Luneburg
(Luneburg, 1966), D-O

nevertheless \ underestimated the nature of connection between the Coulombic
and optical

(fish-eye) problems described in the book by Luneburg. The assumption of only

"close relationship" caused D-O to replace Eq.(3) by

\begin{equation}
\lbrack -\frac{\hslash ^{2}}{2m}\nabla _{i}^{2}+V_{eff}(\mathbf{r}_{i})]\psi
(\mathbf{r}_{i})=0.  \tag{4}
\end{equation}

Eq.(4) is looking differently from Eq.(3). Eq.(3) is an eigenvalue spectral
problem while

Eq.(4) is the Sturmian problem. That is to say, for the Sturmian-type
problem to be well

defined, the parameters entering into $V_{eff}(\mathbf{r}_{i})$ must be
quantized. Such quantization of

parameters is making Sturmian and eigenvalue problems equivalent. This
circumstance

is nontrivial and requires some explanations and examples. These were
provided in

(Kholodenko and Kauffman, 2019) but were entirely overlooked by D-O.
Accordingly,

they were also overlooked in (Thussen and Ceulemans,

2017)\footnote{%
E.g. read above in "Works by Demkov and Ostrovsky" subsection.}. Overlooking
this circumstance is excusable since it is caused by the gaps

in mathematical physics education of physicists and chemists\footnote{%
E.g. read "Summary of solved problems.General background"}.To correct this

problem, that is to provide missing details, our readers are encouraged to
look into

(Kholodenko and Kauffman, 2019). Surprisingly, from this work our readers

will find that in many instances use of Eq.(4) is more advantageous than use
of Eq.(3).

In (Thussen and Ceulemans, 2017), page 377, \ we read that Eq.(4) "does not
describe the

bound states of the atom".That this is not the case could be concluded
already by D-O

themselves should they read the corresponding \ places in books by Lunenburg

(Luneburg,1966) and Caratheodory (Caratheodory, 1937). Ironically, D-O do

quote both of these references in (Demkov and Ostrovsky, 1971). This
misunderstanding

of importance of Eq.(4) and its relation with Eq.(3) resulted in subsequent
critique

and neglect of D-O works, e.g. read (Thussen and Ceulemans, 2017), page 377.
\ 

In (Kholodenko and Kauffman, 2019) and papers which will follow later we

demonstrate that Maxwell's fish-eye and related to the fish-eye classical
and quantum

problems are not \textbf{closely related to} the Coulombic (hydrogen atom)
problems as stated

by D-O. Instead, the Maxwell fish-eye problem is \textbf{isomorphic }to the
Coulombic problem

both classically and quantum mechanically. By overlooking the
Coulombic-fish-eye

isomoprphism at the quantum level D-O argued, nevertheless, that Schr\"{o}%
dinger's

Eq.(3) with Coulombic and fish-eye-type potentials \textbf{both} possess the
O(4,2)

(or SO(4,2)) dynamical symmetry known for the hydrogen atom (Englefield,1972)

and later established for the rest of atoms of periodic table, e.g read
(Thussen and

Ceulemans, 2017)and references therein. Use of the fish-eye-like potentials

by D-O \ was guided in part by their desire to describe the \textsl{atoms} 
\textsl{other than}

\textsl{hydrogen\footnote{%
Since, as we explained already, D-O believed that by using the fish-eye
potential instead of Coulombic it will become possible to overcome
limitations of classical Bertrand theorem applied to multielectron atoms.}. }%
In addition, though, they believed that: a)

$V_{eff}(\mathbf{r}_{i})$ in Eq.(4) \ can be represented by the conformally
deformed fish-eye potential

because, unlike Eq.(3), Eq.(4) is manifestly conformally invariant\footnote{%
Details are explained in (Kholodenko and Kauffman, 2019)} so that,

based on arguments presented above, b) use of such (conformally deformed)

potential removes restrictions posed by the Bertrand theorem in flat space.

His paper \ (Ostrovsky, 1981) Ostrovsky concludes with the following remark:

"It \footnote{%
...that is the group-theoretical (our insert from previous discussion)
consideration...} leaves a very interesting question unresolved, the
question of why the

interaction of a number of electrons with each other and with an atomic
nucleus leads

to an effective one-electron potential having some approximate hidden
symmetry.

The method of solution of this question can hardly be envisaged at the
present time."

Thus, with formal success of quantum mechanical description of atoms of the
whole

periodic system culminating in the formal proof of Madelung rule, Ostrovsky
admits

that with all his results published to date, the L\"{o}wdin's challenge
problem still remains

out of reach. This is so because the deformed fish -eye potential used in
D-O calculations

had no visible connection with the $V_{eff}(\mathbf{r}_{i})$ coming from the
Hartree-Fock calculations.

That is D-O were unaware of such a connection. In addition \ to their
inability to solve

the L\"{o}wdin challenge problem, D-O also failed to solve the \textsl{%
Bertrand challenge\footnote{%
Our observation}}:

\ \ \ What makes the deformed fish-eye potential used by D-O as substitute
of $V_{eff}(\mathbf{r}_{i})$

so good that it removes the restrictions of the classical Bertrand
theorem?\bigskip

\textbf{Analytical equivalence of the Hartree-Fock} $V_{eff}(\mathbf{r}_{i})$
\textbf{\ and the deformed }

\textbf{fish-eye potential\bigskip }. \textbf{The place of \ Bertrand
spacetimes in this equivalence\medskip }

In their cycle of works on proving the Madelung rule D-O used the fish-eye ($%
\gamma =1)$ and

conformally deformed ($\gamma \neq 1)$ fish -eye potentials%
\begin{equation}
V(x,y,z)=-\left( \frac{a}{r}\right) ^{2}\left[ \frac{n_{0}}{\left(
r/a\right) ^{-\gamma }+\left( r/a\right) ^{\gamma }}\right] ^{2},  \tag{5}
\end{equation}

$r^{2}=x^{2}+y^{2}+z^{2},a=const,\gamma $ is a rational number, as an
alternative to the $V_{eff}(\mathbf{r}_{i})$

Hartree-Fock type potentials routinely used in atomic physics literature.
Such a

replacement required them to switch from Eq.(3) to the conformally invariant
Eq.(4)

for reasons just mentioned above. More details are provided in (Kholodenko
and

Kauffman, 2019). Since Eq.(4) seemingly allows only to look for
eigenfunctions

with zero eigenvalue, both D-O and the rest of researchers in the field
considered

this limitation as serious deficiency. Based on results of (Kholodenko and

Kauffman, 2019) summarized in previous subsection we claim that, on the
contrary,

this restriction is harmless and, in fact, very helpful. Such a replacement
of Eq.(3) by

Eq.(4) was made by D-O for the purpose of taking care of limitations of the
classical

Bertrand theorem. No other authors, including those performing Hartree-Fock

calculations, were concerned with these limitations. In the case of
Hartree-Fock type

calculations this lack of concern superimpoosed with the fact that $V_{eff}(%
\mathbf{r}_{i})$ is centrally

symmetric created a serious problem of deriving and describing semiclassical
(and,

hence, classical) limit of quantum multielectron models of atoms other than
hydrogen.

Use of Eq.(4) with potential (5) allowed D-O to neglect works by other
authors on

the same or related subjects and to draw attention of others to their own
works. This

happens to be a fundamental drawback causing D-O to acknowledge that, in
spite of

all their efforts, they still failed to solve the L\"{o}wdin problem. D-O
realized that

when the potential, Eq.(5), used in Eq.(4) the constant $n_{0}$ must acquire
discrete values

as it happens in all Sturmian type problems. Furthermore, for $\gamma =1/2$
the solution of

Eq.(4) provides results compatible with the Madelung rule. The apparent
limitation,

$E=0,$ along with \ no apparent relationship between the potential $V(x,y,z)$
and $V_{eff}(\mathbf{r})$

coming from the Hartree-Fock calculations caused Ostrovski to acknowledge

(Ostrovsky, 1981) that all D-O results to date do not solve the L\"{o}wdin
challenge problem.

Thus, we are left with the following facts:

a) use of the potential, Eq.(5), apparently removes the limitations of the
classical

\ \ \ \ Bertrand theorem;

b) the choice $\gamma =1/2$ in Eq.(5) apparently consistent with the
empirically observed

\ \ \ \ Madelung rule;

c) based on the existing mathematical background of physics and chemistry
community,

\ \ \ \ \ finding of the spectral results beyond $E=0$ requires use of
sophisticated perturbational

\ \ \ \ methods described in D-O works;

d) the choice $\gamma =1/2$ in Eq.(5) is completely detached from known
Hartree-Fock

\ \ \ \ results for $V_{eff}(\mathbf{r})$;

e) the case $\gamma =1$ corresponds to the standard Maxwell's fish-eye
potential. \ 

\ \ \ \ Classical dynamics in such a potential is isomorphic to that in the
Kepler-Coulomb

\ \ \ \ potential, that is for the Bohr-Sommerfeld model of hydrogen atom.
However,

\ \ \ \ because of the apparent $E=0$ limitation at the quantum level,
neither D-O nor

\ \ \ \ other researchers \ reproduced known eigenvalue spectrum for the
hydrogen atom

\ \ \ \ using Eq.(4).

Subsequently, other authors studied \ Eq.(4) with D-O potential, Eq.(2), in
2 dimensions

where use of conformal transformations leaves Eq.(4) form-invariant. In 3
dimensions

one has to use more sophisticated methods of treatments of conformal
transformations.

These are described in great detail in (Kholodenko and Kauffman, 2019). Form

invariance of two dimensional results provides many technical advantages. In
spite of this,

no attempts to reproduce known 2 dimensional results for hydrogen atom were
made

till the work by Kholodenko and Kauffman. In (Kholodenko and Kauffmanm 2019)
we

use the observation ( in section 4) that results on \textbf{R}$^{2}$ can be
lifted to $S^{2}$

and then lifted further to $S^{3}$ via Hopf mapping. Basic facts on Hopf
mapping can be

found either in our book (Kholodenko, 2013) or, in condensed form, in
(Kholodenko

and Kauffman, 2019). Using stereographic projection: from $S^{3}$ to \textbf{%
R}$^{3}$, it is possible

then to reobtain the D-O results done on \textbf{R}$^{3}$. Even though the
connection between

the Hartree-Fock and the D-O potential, Eq.(5), will be discussed in detail
from

geometrical and topological perspective in later works, already described
results

allow us to discuss rigorously some aspects of such a connection now.

Going back to a), we direct our readers attention to the work by Volker
Perlick

(Perlick, 1992). In it results of the classical Bertrand theorem (Goldstein
et al,

2014) valid in Euclidean 3 space had been generalized to static

spherically symmetric spacetimes of general relativity. By design, the
motion in

such curved spacetimes takes place on closed orbits. Detailed calculations

performed in (Kholodenko and Kauffman, 2019) demonstrate that the potential,

Eq.(5), indeed, removes the limitations of the classical Bertrand theorem
since it is

actually working not in flat Euclidean \textbf{R}$^{3}$ but in curved
Bertrand spacetime.

The choice $\gamma =1/2$ listed in b) and d) is indeed connected directly

with results of Hartree-Fock calculations and with Madelung rule. In atomic

physics literature the potential, Eq.(5), $\gamma =1/2,$ is known as the
Tietz potential.

It is bearing the name of his creator. Its origin and many properties are
discussed

in the book by Flugge (Flugge, 1999). Its remarkable numerical coincidence
with

the Hartree-Fock type potential $V_{eff}(\mathbf{r}_{i})$ was discussed in
many places, e.g. read

(Kirzhnitz et al,1985), p.664, Fig.10. Tietz, the author who invented the
Tietz

potential, was initially driven by the desire to simplify the Thomas -Fermi
(T-F)

calculations. \ Much more analytically cumbersome T-F type potentials were
used

by Latter (Latter, 1955) in his numerical study of Schr\"{o}dinger's
equation spectra

of \ low lying excitations for all atoms of periodic system. The numerical
results of

Latter had been subsequently analyzed by March. On page 76 of his book

(March, 1975), without explicit mention of the Madelung rule, March described

results by Latter in terms of the Madelung rule.

After discovery of the potential, now bearing his name, Tietz used it in the

stationary Schr\"{o}dinger equation, Eq.(3), in which $V_{eff}(\mathbf{r}%
_{i})$ was replaced by

the Tietz potential, that is by Eq.(5) with $\gamma =1/2$ (Tietz,
1956).Tietz used

Eq.(3) in which $E\neq 0.$ This is in striking departure from the D-O
version of

this equation, that is Eq.(4), in which $E=0$ by design. In the light of
results

of Appendix F of (Kholodenko and Kauffman, 2019) this happens to be

permissible. It is exactly this fact which makes our calculations different
from

any other performed by standartly trained physical chemists. Unlike Tietz
and, in

accord with D-O, we used Eq.(4) for solving the corresponding eigenvalue

problem. Our method of solving this equation differs from that used by D-O.

The fundamental drawback of D-O method of solving Eq.(4) lies in its

inability to reproduce the classical hydrogen atom

spectrum (problem e)). Therefore, it cannot be considered as reliable and

mathematically sound. At the same time, we had began our study of solutions

of Eq.(4) by using potential, Eq.(5), with $\gamma =1.$ We succeeded in
developing

new method of reproducing hydrogen atom spectrum in 3 dimensions. In

addition, we reproduced this spectrum correctly for the 2 dimensional version

of the quantum hydrogen atom model as well. Although our method

differs from that proposed by Fock in 1936 (Singer, 2005), there is some

overlap to be discussed in the next section. Developed method allowed

us to bypass entirely the most cumbersome item c) present in D-O works.

After solving Eq.(4) with potential, Eq.(5), $\gamma =1,$ correctly, we
obtained

the low lying spectrum for any atom of the periodic system of elements by

employing Eq.(4) with potential, Eq.(5), $\gamma =1/2.$ The obtained results
are

consistent with the empirical Madelung rule.

Incidentally, Tietz also recognized that Eq.(3) with his potential, that is

Eq.(5), $\gamma =1/2,$ can be solved exactly. His first attempt to do exact

calculation was made in 1956 (Tietz, 1956).The rest of his attempts is

summarized in (Tietz, 1968). Obtained exact solutions differ substantially

from those used in D-O works. Besides, Tietz (Tietz, 1956) used his exact

solution only to check it against known Hartree-Fock results for the

Mercury ($Z=80$). In doing so he got a very

good agreement with published results but never tried to extend the

comparison of his exact results for other Z's with those obtained by

the Hartree-Fock methods.\bigskip

{\Large Beyond the canonical Madelung rule\bigskip }

\textbf{General comments}

\bigskip

With accuracy of the existing Hartree-Fock methods, including their
relativistic

versions (Dyall and Faegri, 2007), the question arises: Why one should care

about the Madelung rule and its exceptions? To answer this question, we would

like to mention the fact, noticed initially by Symanzik (Symanzik, 1966),
that

all quantum field theories used in particle physics, e.g. those used for

description of the Standard Model-an analog of the periodic system at the
level of

elementary particles, are describable in terms of the models of polymer
chains

used in polymer physics. More on this is given in the paper by Aisenman

(Aisenman,1985), and references therein, significantly developing Symanzik's

ideas. The moral of these studies is simple:

instead of building expensive particle accelerators to study physics of
elementary

particles, it is sufficient to study properties of polymer solutions in the
lab.

Initially, the (quark model, quark symmetry) ideas of particle physics were
used

for classification of elements in the periodic system of elements by Fet
(Fet, 2016).

These were also discussed in some detail in (Thussen and Ceulemans, 2017)
and by

(Varlamov, 2018) without emphasis on the Madelung rule though. The task \ of
this

and future publications is to demonstrate that the noticed fruitful
cross-fertilization

between just mentioned results of physics and chemistry might potentially
yield

new and significant results both in physics and chemistry if the exceptions
to the

Madelung rule in the atomic physics are to be studied\medskip . Below, we
initiate such

cross -fertilization process with the following observation\medskip

\textbf{Quantum defects from methods of general relativity\bigskip\ and
Dirac equation}

In 1890 Ridberg conjectured (and tested) that for multielectron atoms the
energy

spectrum may be written in the form $-\frac{1}{2n^{2}}$ resembling that for
hydrogen atom

(Burkhardt and Leventhal, 2006). Since the accidental degeneracy is
nonexistent

for multielectron atoms the energy is a function of both the principal
quantum

number $n$ and $l$ is the angular momentum quantum number. Specifically, in
the

appropriate system of units the energy spectrum for multielectron atoms \
can be

presented as%
\begin{equation}
E_{n,l}=-\frac{1}{2(n-\delta _{l})^{2}}.  \tag{6}
\end{equation}

This formula defines the quantum defect $\delta _{l}.$ (Burkhardt et al,
1992) demonstrated

that calculation of $\delta _{l}$ can be accomplished in exactly the same
way as calculation

of the perihelion shift for the Mercury in Einsteinian theory of relativity.
In such a

case $\delta _{l}$ is proportional to this shift. This fact hints that in
the case of multielectron

atoms relativistic effects may play an important role. \ Since in

(Kholodenko and Kauffman, 2019) the Madelung rule was obtained within a scope

of many-body nonrelativistic quantum mechanics, not surprisingly, a
comparison

with experimental data indicates that this rule is not universal across the
periodic

table even though it works rather well for the majority

of elements to the extent that some authors, e.g. read chapter 5 of (Scerri
and

Restrepo, 2018) insisted that, if properly interpreted, the Madelung rule is

applicable for the whole periodic system of elements. This fact is supported
by

the group-theoretic considerations leading to the conclusion that the
underlying

symmetry of the periodic table is SO(4,2) and, using this symmetry, that the

Madelung rule(without exceptions) \ follows from the group theoretic

considerations (Fet, 2016), (Thussen and Ceulemans, 2017), (Varlamov, 2018).

Experimental data\footnote{%
https://en.wikipedia.org/wiki/Aufbau\_principle} indicate, nevertheless,
that the Madelung rule

does have exceptions. All of them are coming from the heavier elements of

periodic table. In fact, it is surprising that about 2/3 of elements of the
periodic

table does obey the Madelung rule in its canonical form\footnote{%
Stated above, in this work.}.

Thus, we are faced with the problem of explaining why at least 2/3 of
elements

do obey the canonical Madelung rule and what mechanisms break this rule.

This problem can be alternatively restated as follows. Why for the most
elements

of periodic table the relativistic effects are negligible and, why without
exceptions,

they are significant in the case of elements exhibiting the Madelung

rule anomalies? \ 

As we just mentioned, theory of quantum defects should, in principle, provide

needed answer. But mentioned results based on relativistic calculations known

from the theory of theory of perihelion shift are not the only ones which
can be used

as point of departure. Another approach of computation of quantum defects
coming

very close to ideas and methods developed in (Kholodenko and Kauffman, 2019)

is presented in (Karwowski and Martin,1991) and later, in (Martin, 1997).
Both papers

use Dirac equation in second order form as point of departure. In its second
order form

this equation differs very little from the nonrelativistic Schr\"{o}dinger
equation treated

in (Kholodenko and Kauffman, 2019). Although not mentioned in just cited
works,

below we shall argue that calculations based on the perihelion shift
(Burkhardt et

al, 1992) and on use of the Dirac equation (Martin, 1997) are physically not
too

much different from each other. This fact is of fundamental significance. We
shall

fully describe it elsewhere.\medskip

\textbf{Phenomenology of the canonical Madelung rule \bigskip }

In previous subsection we noticed that calculations of quantum defect

$\delta _{l}$ proceed in complete analogy with those for the perihelion
shift in general relativity.

This fact indicates that many quantum mechanical features can be

explained (visualized) with help of macroscopic phenomena. This was realized

already by Darwin shortly after invention of quantum mechanics, e.g. read
his book

(Darwin, 1931), chapter 5. Darwin used the theory of standing waves (e.g. on
the

vibrating string) and extended it to two and three dimensions. He did this
with the

concept of a node.

\textsl{A} \textsl{node} \textsl{is a point on the string which does not
move during the }

\textsl{vibration.}

\textsl{\ }When going to two dimensions, nodes are no longer points but
lines. E.g.

nodal modes of a drum made in a shape of the disc are either radial lines
through the

disc center or the set of concentric circles around the disc center. Darwin
noticed

that: "a \ quick and easy way of describing the various modes, is by taking
two

numbers, the first of which stands for the number of circular nodal lines
and the

second for the number of straight radial ones." \ In the case of three
dimensions

"\textsl{we shall get nodal surfaces instead of nodal lines}. \textsl{These
may be either spheres,}

\textsl{\ or else planes or perhaps cones through the centre."} Just
described nodal patterns

Darwin connects with the nodal patterns of wave functions for hydrogen atom.
His

results had been discussed further by Born (Born, 1936), chapter 4. In his
book he

mentions about \textsl{Chladni's figures}. More details/references on these
figures are

given in (Kholodenko, 2017).These figures can be readily visualized in the
case of

a circular drum. For this, it is sufficient to cover drum with sand and to
make it vibrate.

The sand will remain only at the places where there is no vibration. By
definition, the

fundamental tone exhibits no nodal lines. If these devices are to be
compared with the

nodal patters of, say, hydrogenic wave functions, the non-vibrating boundary
should

be counted as nodal line. This then resolves the apparent difficulty: In the
atom, there

is no fixed boundary. Instead, there is an atomic nucleus attracting the
electron.

Wiswesser extended this single electron picture to the multielectron atoms.
He also

accounted to the Pauli rule in (Wiswesser,1945). Completely independently
such a

generalization was made in (Steen et al, 2019). In this work all possible
droplet

motions/vibrations were classified taking into account surface tension
acting on the

deformable droplet surface. Although Wiswesser was apparently

unfamiliar with the Madelung rule, his way of analyzing nodal patterns of
different atoms

had lead him to conclude that the nodal patterns consistent with aufbau
filing

are possible only if the Madelung rules holding. On page 319 (bottom) of
(Wiswesser,1945)

we find the statement: "the patterns will be filled in increasing order of 
\textit{n+l." }

Here $n$ and $l$ were defined in the previous subsection. More specifically,
following

Darwin's logic, we \ notice that:

1. Any 3 dimensional nodal pattern is being characterized by 3 numbers $a,b$
and $c$.

2. Any atomic wave function is characterized by at least 3 quantum numbers $%
n,l$ and $m.$

3.The relationship between the numbers $a,b$ and $c$ and $n,l$ and $m$ is
given as follows:

$n=a+b+1$, $l=b$, $m$ may take all integer values between $\pm b.$ To extend
these rules to

multielectron atoms Wiswesser accounted for the spin. His results are
summarized in the

Table1 of his paper.

Unlike (Wiswesser,1945), the results of (Steen et al, 2019) indicate that:

a) The Madelung-type filling is taking place when \textit{n+l=even;}

b) With increasing $n,$ the Madelung rule becomes irregular.

It is completely useless for us to identify the observed irregularities in
the

vibrational patterns of fluid droplets with the exceptions to the validity
of the Madelung

rule in periodic system. In the next subsection we shall begin explanation
why this is so.

In the reminder of this section additional details will be added.\medskip

\textbf{The canonical Madelung rule obtained microscopically and its
relativization\medskip }

Since the canonical Madelung rule was obtained microscopically in subsection
4.6 of

(Kholodenko and Kauffman, 2019) there is no need to repeat the derivation
here.

Nevertheless, in this work it is of interest to connect this derivation with
phenomenological

results of \ previous section. This connection will also help us to develop
the formalism

in such a way, that it \ will become possible to treat the exceptions to the
canonical

Madelung rule. \ In doing so we shall initially follow (Kholodenko and
Kauffman, 2019),

(Englefield,1972) and (Biedenharn and Louck, 1981).

Specifically, in a specially chosen system of units in \ which the
Hamiltonian H for hydrogen

atom is dimensionless, it is given in the operator form by%
\begin{equation}
\text{\^{H}}=\mathbf{p}^{2}-\frac{2}{r}  \tag{7}
\end{equation}

, the Laplace-Runge-Lenz vector \textbf{A}$_{0}$ is given by%
\begin{equation}
\mathbf{A}_{0}=\frac{\mathbf{x}}{r}+\frac{1}{2}(\mathbf{L\times p-p\times L)}
\tag{8}
\end{equation}

$\mathbf{,}$while the angular momentum operator \textbf{L} is defined as
usual by \textbf{L}=\textbf{x}$\times \mathbf{p.}$ It is convenient

to normalize \textbf{A}$_{0}$ as follows%
\begin{equation}
\mathbf{A}=\left\{ 
\begin{array}{c}
\mathbf{A}_{0}(-H)^{\frac{1}{2}}\text{ for E\TEXTsymbol{<}0,} \\ 
\mathbf{A}_{0}\text{ for E=0,} \\ 
\mathbf{A}_{0}=(H)^{\frac{1}{2}},\text{ for E\TEXTsymbol{>}0.}%
\end{array}%
\right.  \tag{9}
\end{equation}

Here it is assumed that \^{H}$\Psi _{E}=E\Psi _{E}$ and $E=H.$ By
introducing two auxiliary angular

momenta \textbf{J}($\alpha ),\alpha =1,2,$ such that \ \textbf{J}($1)=\frac{1%
}{2}(\mathbf{L}+\mathbf{A})$ and \textbf{J}($2)=\frac{1}{2}(\mathbf{L}-%
\mathbf{A}),$ and using

known commutation relations for \textbf{L}, etc., we arrive at 
\begin{eqnarray}
\mathbf{J}(\alpha )\times \mathbf{J}(\alpha ) &=&i\mathbf{J}(\alpha ),\alpha
=1,2  \TCItag{10} \\
\lbrack \mathbf{J}(1),\mathbf{J}(2)] &=&0  \notag
\end{eqnarray}

Taking into account that \textbf{L}$\cdot $\textbf{A}=0 we also obtain two
Casimir operators: \textbf{L}$\cdot $\textbf{A}=0=\textbf{A}$\cdot \mathbf{L}
$ and

$\mathbf{L}^{2}+\mathbf{A}^{2}$. The Lie algebras $\mathbf{J}(\alpha )\times 
\mathbf{J}(\alpha )=i\mathbf{J}(\alpha ),\alpha =1,2,$ are the algebras of
rigid rotators

for which the eigenvalues $j_{\alpha }(j_{\alpha }+1)$ are known from the
standard texts on quantum

mechanics. The peculiarity of the present case lies in the fact that $%
\mathbf{J}(1)^{2}=\mathbf{J}(2)^{2}$. This

constraint is leading to the requirement: $j_{\alpha }=j_{\beta }=j.$ The
topological meaning of this

requirement is explained in subsection 5.3.3. of \ (Kholodenko and Kauffman,
2019).

In short, the eigenvalue equation for the standard quantum mechanical rigid
rotator

is that for the Laplacian living on $S^{2}$. \ Since in the present case we
are having two

rigid rotators, each of them should have its own sphere $S^{2}$. However,
the constraint

$j_{\alpha }=j_{\beta }=j$ causes these two spheres to be identified with
each other pointwise.

Topologically, such a poinwise identification leads to the sphere $S^{3}.$
Group-theoretically

the same result can be stated as $so(4)\simeq so(3)\oplus so(3).$

With such background we are ready a) to connect the results of previous
subsection

with those just defined and b) to relativize these results.

We begin with the first task. We proceed by known analogy. The 3 dimensional
rigid

rotator eigenvalues and eigenfunctions are solutions of the equation%
\begin{equation}
\mathbf{L}^{2}Y_{lm}(\theta ,\phi )=l(l+1)Y_{lm}(\theta ,\phi ).  \tag{11}
\end{equation}

However \textbf{L}$^{2}=$L$_{x}^{2}+$L$_{y}^{2}+$L$_{z}^{2}$ and L$_{x}=i$D$%
_{23},$L$_{y}=i$D$_{31},$L$_{z}=i$D$_{12}$ , where 
\begin{equation}
D_{\alpha \beta }=-x_{\alpha }\frac{\partial }{\partial x_{\beta }}+x_{\beta
}\frac{\partial }{\partial x_{\alpha }},\text{ \ \ \ }\alpha <\beta =1,2,...d
\tag{12}
\end{equation}

where d is the dimensionality of space. In 4 dimensions, following
(Englefield,1972) we

can put A$_{x}=iD_{14},$A$_{y}=iD_{24},$A$_{z}=iD_{34}.$Thus, if $\mathbf{L}%
^{2}$ represents a Laplacian on $S^{2},$the

combination $\mathbf{L}^{2}+\mathbf{A}^{2}\equiv \mathcal{L}^{2}$ represents
a Laplacian \ on $S^{3}.$That is, instead of more familiar

(from standard textbooks on quantum mechanics) study of rigid rotator on
two-sphere,

$S^{2},$the eigenvalue problem for hydrogen atom actually involves study of
spectrum of the

rigid rotator on 3-sphere. The 3 Euler angles $\alpha ,\theta ,\phi $ on the
3 sphere are replacing more

familiar $\theta ,\phi $ angles used for the 2 sphere. The eigenvalue
Eq.(11) now is \ being

replaced by 
\begin{equation}
\mathcal{L}^{2}Y_{nlm}(\alpha ,\theta ,\phi )=I_{nl}Y_{nlm}(\alpha ,\theta
,\phi )  \tag{14a}
\end{equation}

This result is almost ready for comparison with that discussed in the
previous subsection

because in both cases we are having manifestly spherically symmetric wave
functions

\ with indices $n,l,m$. To replace "almost ready" \ with "ready" \ we only
have to notice

that the conformal transformations, e.g. those in Eq.(5), only cause
relabeling of

\ the indices in Eq.(14a), e.g. the choice $\gamma =1$ in Eq.(5) leads to
Eq.(14a) (as required

for hydrogen atom) while the choice $\gamma =1/2$ leads to 
\begin{equation}
\mathcal{L}^{2}Y_{n+l,lm}(\alpha ,\theta ,\phi )=I_{n+l,l}Y_{n+l,lm}(\alpha
,\theta ,\phi )  \tag{14b}
\end{equation}

implying the canonical Madelung rule. In (Wiswesser,1945) the relabeling of
indices $n,m$

and $l$ was not connected with the Hartree-Fock calculations, etc. and,
therefore, cannot

be considered as ab initio derivation of the canonical Madelung rule The ab
initio proof

of this rule is given in (Kholodenko and Kauffman, 2019). Since the
canonical rule has

exceptions, we are now in the position to relativize the obtained results.

This task requires several steps. First, we notice that in standard 3
dimensional

calculations the hydrogen spectrum is determined by the \textsl{radial
equation}%
\begin{equation}
\lbrack -\frac{1}{2}(\frac{d^{2}}{dr^{2}}+\frac{2}{r}\frac{d}{dr}-\frac{%
l(l+1)}{r^{2}})+V(r)]R_{El}(r)=ER_{El}(r).  \tag{15}
\end{equation}

The total wave function $\Psi _{E}=F_{El}(r)\mathcal{Y}_{lm}(\theta ,\phi ),%
\mathcal{Y}_{lm}(\theta ,\phi )=r^{l}Y_{lm}(\theta ,\phi
),R_{El}(r)=r^{l}F_{El}(r)$

and $V(r)=-\frac{Ze^{2}}{r},m=1,\hbar =1.$The combination $F_{El}(r)\mathcal{%
Y}_{lm}(\theta ,\phi )$ can be rewritten in

terms \ of $Y_{nlm}(\alpha ,\theta ,\phi )$ as demonstrated in (Kholodenko
and Kauffman, 2019) and with

help of other references therein. Therefore, it is sufficient to look at 3
dimensional results.

They can always be mapped onto $S^{3}$ using the inverse stereographic
projection.

Next, this observation allows us, following Martin and Glauber (Martin and
Glauber,1958)

and Biedenharn (Biedenharn, 1983), to use the Pauli matrices $\sigma _{i}$
in order to rewrite

\textbf{L}$^{2}=\left( \mathbf{\sigma }\cdot \mathbf{L}\right) \mathbf{%
(\sigma }\cdot \mathbf{L+}1).$ This identity permits then to write the total
momentum \textbf{J} as

$\mathbf{J}=\mathbf{L}+\frac{1}{2}\mathbf{\sigma }$ . It is convenient then
to introduce the operator $\mathcal{K=}\mathbf{\sigma }\cdot \mathbf{L+}1$
introduced

already by Dirac (Dirac,1958). With help of \ this operator it is possible
to obtain an

identity $\mathcal{K}^{2}=\mathbf{J}^{2}+\frac{1}{4},\hbar =1.$The
eigenvalues of $\mathcal{K}$ will be \ denoted by $\kappa $. They are:

$\kappa =\pm 1,\pm 2,...(0$ is excluded$).$ From these definitions it
follows that%
\begin{eqnarray}
l &=&l(\kappa )=\left\{ 
\begin{array}{c}
\kappa ,\text{ if }\kappa \text{ is positive} \\ 
\left\vert \kappa \right\vert -1,\text{ if }\kappa \text{ is negative}%
\end{array}%
\right\vert  \notag \\
j &=&j(\kappa )=\left\vert \kappa \right\vert -\frac{1}{2}.  \TCItag{16}
\end{eqnarray}

The above definitions were made with the purpose not emphasized at all in
standard texts

on quantum mechanics. Specifically, at the classical level Kepler
trajectories can be

determined with help of \textbf{A} only (Collas, 1970).This fact suggests
that the quantum analog

of \textbf{A} should produce the eigenvalue spectrum identical to that
obtained from Eq.(15).

This is indeed the case. To demonstrate this we introduce the operator $%
\mathcal{N}$ such that

$\left( \mathcal{N}\right) ^{2}=\left( \mathbf{\sigma }\cdot \mathbf{A}%
\right) ^{2}$ +$\left( \mathcal{K}\right) ^{2}.$ Since it can be shown that $%
\mathbf{\sigma }\cdot \mathbf{A}$ and $\mathcal{K}$ anticommute, it

becomes also possible to write%
\begin{equation}
\mathcal{N}=\mathbf{\sigma }\cdot \mathbf{A}+\mathcal{K}  \tag{17}
\end{equation}

Denote the eigenvalues of $\mathcal{N}$ as $\pm N.$ Then, it is possible to
demonstrate that 
\begin{equation}
\mathbf{\sigma }\cdot \mathbf{A\mid }N,\varkappa ,m>=(N^{2}-\varkappa ^{2})^{%
\frac{1}{2}}\mid N,-\varkappa ,m>.  \tag{18}
\end{equation}

It is possible to demonstrate that $N\rightleftarrows E$ with $E$ defined in
Eq.(15). With help of this

result it is possible to write an exact equivalent of the radial Eq.(15). It
is given by 
\begin{equation}
\lbrack \frac{1}{r^{2}}\frac{d}{dr}r^{2}\frac{d}{dr}-\frac{\mathcal{K}(%
\mathcal{K}+1)}{r^{2}}+\frac{2Ze^{2}}{r}-k^{2}]F_{N,l(\kappa )}(r)=0. 
\tag{19}
\end{equation}

Here $k^{2}=2\left\vert E\right\vert ,m=1,\hbar =1.$Biedenharn (Biedenharn,
1983) explains how the wave

function $\mid N,-\varkappa ,m>$ \ is related to $F_{N,l(\kappa )}(r).$Also, 
$\mathcal{K}(\mathcal{K}+1)=l(\kappa )(l(\kappa )+1).$ Not only

just presented results demonstrate that the quantum version of the \
Laplace-Runge-Lenz

operator leads to the eigenvalue problem identical to the standard
eigenvalue problem,

Eq.(15), presented in every textbook on quantum mechanics.

In addition, these results permit relativistic generalization. The control
parameter in this

generalization is the fine structure constant $\alpha =\frac{e^{2}}{c\hbar }%
. $ In the limit $\alpha =0$ the result, Eq.(19),

is recovered while for $\alpha >0$ is replaced by very similarly looking
equation\footnote{%
Here, to avoid confusion, when comparing with original sources, we restore $%
\hbar ,c$ and $m$.} 
\begin{equation}
\lbrack \frac{1}{r^{2}}\frac{d}{dr}r^{2}\frac{d}{dr}-\frac{\Gamma (\Gamma +1)%
}{r^{2}}+\frac{2\alpha ZEe^{2}}{c\hbar r}-k^{2}]\Phi _{N,l(\gamma \kappa
)}(r)=0.  \tag{20}
\end{equation}

Here $k^{2}=[\left( m^{2}c^{4}-E^{2}\right) /c^{2}\hbar ^{2}],\Gamma $ is
the Lippmann-Johnson operator%
\begin{equation}
\Gamma =\mathcal{K}+i\alpha Z\rho _{1}\mathbf{\sigma }\cdot \mathbf{\check{r}%
},  \tag{21}
\end{equation}

$\mathbf{\check{r}}=\frac{\mathbf{x}}{r},\rho _{1}\div \rho _{3},\sigma
_{1}\div \sigma _{3}$ are $4\times 4$ matrices defined in Dirac's book
(Dirac,1958). Instead of

eigenvalue $\kappa $ for $\mathcal{K}$ now one has to use $\gamma \kappa $
so that, upon diagonalization, $\Gamma (\Gamma +1)=l(\gamma \kappa
)(l(\gamma \kappa )+1)$

and 
\begin{equation}
l(\gamma \kappa )=\left\{ 
\begin{array}{c}
\gamma \kappa =\left\vert \kappa ^{2}-\left( \alpha Z\right) ^{2}\right\vert
^{\frac{1}{2}}\text{ for }\gamma \kappa >0 \\ 
\left\vert \gamma \kappa \right\vert -1=\left\vert \kappa ^{2}-\left( \alpha
Z\right) ^{2}\right\vert ^{\frac{1}{2}}-1\text{ for }\gamma \kappa <0%
\end{array}%
\right\vert  \tag{22}
\end{equation}

The physical meaning of the factor $\gamma $ is to be explained in the next
subsection.

Mathematically, both Eq.s(19) and (20) are looking the same and, in fact,
their

solution can be reconstructed from the solution of radial eigenvalue Eq.(15)

presented in any book on quantum mechanics. The difference lies only in
redefining

the parameter $l.$ In the nonrelativistic case the combination $l(\kappa
)(l(\kappa )+1)$ is the

same as $l$($l$+1) as required, while in the relativistic case we should
replace

$l$ in it by $l(\gamma \kappa ).$By replacing $l$ in Eq.(14b) by $l(\gamma
\kappa )$ it is immediately clear that

the Madelung rule in its canonical form is no longer valid.\medskip

\textbf{Meaning of the }$\gamma $\textbf{\ factor \medskip }

In previous subsection we provided enough evidence that uses of group theory

for derivation of the Madelung rule are destined to fail because standard
methods

cannot be applied for description of exceptions. We outlined reasons

for this to happen by applying accepted rules of both nonrelativistic and
relativistic

quantum mechanics. In the subsection on quantum defects we mentioned about a

peculiar situation of computation of quantum defects purely classically, by
analogy

with perihelion calculations in general relativity, and purely quantum
mechanically,

by using theory of relativistic quantum defects. In this subsection we shall
describe

some fundamental difficulties in formal uses of the existing apparatus of
quantum

mechanics and general relativity. Atomic physics is the most reliable
experimental

and theoretical domain of study of possible changes to both quantum
mechanics and

relativity. This could be seen by studying work by Sommerfeld on fine
structure of

hydrogen atom.

Sommerfeld wrote his seminal paper on fine structure in 1916 (Granovskii,
2004),

the same year Einstein wrote his seminal work on general relativity. To prove

correctness of his theory Einstein, in particular, calculated perihelion
shift of Mercury

(Roseveare, 1982). \ In writing of his paper Sommerfeld was not driven by
this result.

He wanted to extend his own results on extension of Bohr's theory of
quantized

circular orbits. For this purpose he initially extended Bohr's results to
describe the

elliptic orbits. In Bohr's theory the was only one \ quantum number, $n$.
Sommerfeld

added another two quantum numbers: $l$ and $m$ (Sommerfeld,1934). This was
done

before 1916. But in 1916 he decided to reconsider his calculations to
account for

already known effects of fine structure. For this purpose he used the
Hamiltonian of

the type\footnote{%
Here we follow notations of Sommerfeld.}%
\begin{equation}
W=mc^{2}-m_{0}c^{2}-\frac{Ze^{2}}{r},\text{ }m=\frac{m_{0}}{\sqrt{1-\beta
^{2}}},  \tag{23}
\end{equation}

where $\beta ^{2}=\left( \dfrac{\mathbf{v}}{c}\right) ^{2},m_{0\text{ }}$is
the rest mass of electron, \textbf{v} is its velocity. Mass of the nucleus is

taken to be infinite. \ By writing the constant angular momentum $p_{\varphi
}$ as $p$ and by writing

$s=\frac{1}{r}$ Sommerfeld obtained the equation for electron's trajectory 
\begin{equation}
s(\varphi )=C+A\cos \gamma \varphi  \tag{24}
\end{equation}

where $\gamma ^{2}=1-\frac{p_{0}^{2}}{p^{2}},$ $p_{0}=\frac{Ze^{2}}{c}.$ It
happens, that just introduced $\gamma $ is the same as $\gamma $ in

Eq.(22) (Biedenharn, 1983). This fact is fundamental and requires more
explanation than

given by Biedenharn. \ First, the electron mass $m$ in Eq.(23) takes care of
relativistic

effects but the potential term in Eq.(23) is non relativistic. That is it
does not take into

account the retardation effects. Thus, the modification of the Kepler
problem made by

Sommerfeld only takes care of the mass which is becoming velocity-dependent.
This leads

to two fundamental new effects.

a) The Laplace-Runge-Lenz vector is no longer a constant of motion. This
fact removes

\ \ \ the accidental degeneracy.

b) For $\gamma =1$ the electron trajectory, Eq.(24), is describing Kepler's

\ \ \ \ elliptic orbits (Landau and Lifshitz, 1960). For $\gamma <1$ the
orbit never closes because

\ \ \ of the precession analogous to that calculated by Einstein for
Mercury. Sommerfeld

\ \ \ being aware of Einstein's result, also calculated the perihelion shift
by appropriately

\ \ \ choosing constants in Eq.(23), and obtained $7^{\prime \prime }$ per
century. He compared his result

with $43^{\prime \prime }$\ obtained by Einstein for the Mercury and came to
wrong conclusion

\ that his $7^{\prime \prime }$ result have nothing to do with general
relativity.

But, given that the orbit is not closed, it cannot be quantized! Sommerfeld
did not give

up,however, in his search in obtaining fine structure spectrum for hydrogen.
He noticed

that the prihelion shift $\Delta \varphi $ is obtainable as follows: $\Delta
\varphi =\frac{2\pi }{\gamma }-2\pi .$ To cope with this

shift, Sommerfeld transferred his calculations to the rotating system of
coordinates.

For this purpose he introduced angle $\psi =\gamma \varphi $ so that
eventually the orbit became closed.

Application of the Bohr-Sommerfeld quantization prescription had lead him to
the fine

structure spectrum%
\begin{equation}
E(n_{r},n_{\varphi })=m_{0}c^{2}(1+\frac{\alpha ^{2}}{\omega ^{2}})^{-\frac{1%
}{2}},  \tag{25}
\end{equation}

where $\omega =n_{r}+\sqrt{n_{\varphi }^{2}-\alpha ^{2}}.$ Exactly the same
result\footnote{%
Provided that the meaning of n$_{r}$ and n$_{\varphi }$ is slightly
redefined (to account for electron spin)} was later reobtained by Dirac

(Dirac, 1958) with help of Dirac equation. The spectrum was actually
obtained by Darwin

and Gordon in 1928 (Granovskii, 2004). Neither Dirac nor Sommerfeld had
further

investigated the remarkable coincidence between Bohr-Sommerfeld-style
calculations of

the spectrum and truly quantum mechanical claculations. This issue was
addressed in

(Biedenharn, 1983) and, more recently, in (Keppeler, 2003). \ Fortunately,
these people,

have not at all exhausted the topic of this remarkable coincidence.\medskip

\bigskip

\textbf{Challenges for quantum mechanics coming from the effects of
gravity\medskip }

\textbf{a)} \textbf{Motion on a cone}

\textbf{\medskip }

The result $7^{\prime \prime }$ per century for relativistic electron in
Coulomb (or gravitational) field was

obtained already in the work by Poincar$e^{\prime }$ in 1905. This is
described in detail in (Provost

and Bracco, 2018), immediately after their Eq.(11). Poincar$e^{\prime }$,
obtained his result within

a scope of relativistic theory of gravity which he developed. Poincar$%
e^{\prime }$ died in 1912, while

the theory of general relativity inaugurated in 1916 provided $43^{\prime
\prime }.$This was one of its major

hallmarks. Accordingly, for almost a century the perihelion problem was
considered as

closed and Poincar$e^{\prime }$ gravity results were forgotten. The
difficulties with other

formulations, \ e.g. based on Eq.(24) emerged relatively recently. These are
noticed for

the first time here, in this work. Specifically, in (Al-Hashimi and Wiese,
2008) the authors

considered both classical and quantum mechanical dynamics of \ a particle
moving on

a cone and bound to its tip by 1/r potential. At the classical level the
authors obtained

for particle trajectory result identical to our Eq.(24). No attempts to
connect this result

with that obtained by Sommerfeld was made. Subsequently, the results

(Al-Hashimi and Wiese, 2008) had inspired another paper (Brihaye et al,
2014) in which

the same problem was studied from the point of view of the validity of the
Bertrand

theorem. Recall, that the standard Bertrand theorem was obtained for the flat

space. \ The authors intended to extend \ the validity of this theorem to
the conical space.

Again, no reference to the work by Sommerfeld was made. By attempting to
solve the

quantum Bertand problem group-theoretically these authors arrived at the
trivial result:

$\gamma $ in Eq.(24) should be an integer for closed trajectories. If the
problem is studied

traditionally, that is by using the Schr\"{o}dinger equation, then for the
trajectories to be

closed $\gamma $ must be rational. This conclusion was reached in both
papers. Further studies

of the obtained results indicated that the obtained wave functions exhibit
some weird

analytical behavior. Not surprisingly, the obtained spectrum drastically
differs

from that given in Eq.(25). The conical singularities are, in fact,
singularities of

space-time and their presence makes space-time curved (Kholodenko, 2000),

(Al-Hashimi and Wiese, 2008), (Brihaye et al, 2014). Our readers at this
point can

raise an objection.The Hamiltonian,Eq.(23) involves the Coulombic-type
potential.

Although analytically both the gravitational and Coulombic potentials look
the same,

the existing theory of gravity seemingly deals only with modifications of
space-time

caused by the effects of gravity. This is not the case, however, as it was
demonstrated

by Rainich (Rainich,1925) and developed subsequently by Misner and Wheeler in

a form of geometrodynamics (Misner and Wheeler, 1957). In its original form
it

had a problem of including spin into theory. Subsequently, the problem was

resolved. We included this information, just to demonstrate below, that it
is related

to our major task of description of the Madelung rule and its
anomalies.\medskip

\textbf{b) Problems with special relativity in the light of
Sommerfeld\medskip\ results}

According to (Misner et al, 1973), chapter 7, Einsteinian special relativity
is incompatible

with his general relativity. But we just demonstrated that the dynamics
involving special

relativistic Hamiltonian, Eq.(23), is describing, in fact, dynamics typical
for general

relativity. The motion originating with help of Eq.(24) is not inertial as \
required by

the rules of special relativity. In the canonical special relativity only
frames moving

with constant speed are allowed. The motion is taking place with
acceleration. And, if

this is so, then the form of Hamiltonian, Eq.(24), is questionable (let
alone questionable

the instantaneity of the Coulombic interaction).

The difficulty is not removed by Sommerfeld's ingenious trick of
transforming the

problem into rotating system of coordinates. This transition is essential
for quantization

but leaves \ just described problem unsolved. This is so because use of
rotating system

of coordinates is connected with many issues. At the elementary level the
issues are

well described by Diecks in (Rizzi and Ruggiero, 2004), chapter 2. The
bottom line

is the following: as soon as we are in the rotating frame we have to deal
with the

equivalence principle of general relativity. The use of equivalence principle

alone leads us back to the Einsteinian way of calculating the perihelion
shift

(Roseveare, 1982), chapter 7.7. This then creates a problem with\
Sommerfeld's

calculations of a perihelion shift. Use of rotational system of coordinates
is also

associated with the Mach principle (Baratini and Christillin, 2012), (Essen,
2013).

Thus, we are coming to the conclusion:\smallskip

\textsl{Since the Dirac equation is reducible to the Schr\"{o}dinger
equation, it is permissible to}

\textsl{use the Dirac equation in both relativistic and nonrelativistic
calculations\footnote{%
This statement will be elaborated futrther below}.}

\textsl{And if this is so, then the Mach principle of general relativity is
essential}

\textsl{for quantization of (semi) classical orbits}.\smallskip \medskip

That is, contrary to the existing opinions, quantum mechanics and general
relativity are

inseparable. Influence of effects of general relativity (e.g. post-Newtonian
approximation)

which begins with the Hamiltonian, Eq.(24), was initiated in the work by
Kennedy

(Kennedy, 1972). It involves systematic derivation of relativistic
corrections routinely

used in the relativistic quantum many-body calculations (Dial and Faegri,
2007). These

results should be kept in mind when one is thinking about them from the
standpoint of a

remarkable agreement between \ results of Sommerfeld and Dirac for the fine
structure

spectrum. Just mentioned technical difficulties were recognized by Logunov

(Logunov,1989) who noticed that the space-time length interval of special
relativity is

invariant not only with respect to the standard Lorentz transformations used
for inertial

frames known from any textbook on special relativity but with respect to a
much wider

set of transformations characteristic for noninertial frames. The question
then emerges:

what to do with the existing theory of gravitation? Logunov decided to
develop new

theory of gravitation in which gravity is acting very much like the Maxwell's

electrodynamics. Since theory of electrodynamics works perfectly well in
Minkovski

spacetime, apparently, new theory of gravitation might be also working well
in

Minkowski spacetime. \ Although Logunov was able to bring such a theory of
gravitation

to completion, it was not embraced by others for reasons to be explained.
Before

explaining, we would like to mention that, in addition to the book by (Rizzi
and

Ruggiero, 2004), \ recently, there appeared two other books by Lusanna
(Lusanna,

2019) and Gorgoulhon (Gorgoulhon, 2013) discussing special relativity in
general frames.

\ Unlike Logunov, these authors \ stopped short of abandoning the theory of
relativity

in its classical form. The latest developments in general relativity (to be
briefly

sketched below) made these books useful only from the historical
perspective.\medskip

\textbf{c) Madelung rule and its anomalies explained with help of Schr\"{o}%
dinger's}

\ \ \ \textbf{\ work on Dirac electron in a gravitational field\medskip }

The latest developments in general relativity (Aldrovandi and Pereira, 2013)
makes theory

of \ teleparallel gravity the most useful.\ Technically this theory
resembles very much

Yang-Mills theory whose Abelian and non-Abelian versions are being used for
description

of all other types of fields. The idea of teleparallel gravity could be
traced back,

for instance, to 1932\footnote{%
Actually, it was initiated by Einstein himself a bit earlier but Schr\"{o}%
dinger's paper provided some insentive for experimental verification.}. In
this year the paper by Schr\"{o}dinger (Schr\"{o}dinger,

1932) on Dirac electron in gravitational field was published. Historically,
Dirac came

up with his equation (Dirac, 1958) being driven by the observation that the
standard

Schr\"{o}dinger equation is not \ Lorentz invariant. Dirac's equation
corrects this deficiency.

By correcting this deficiency Dirac uncovered spin in 1928. Before, it was
artificially

inserted into Schr\"{o}dinger's equation. The idea for doing \ so belongs by
Pauli. Schr\"{o}dinger

immediately got interested in Dirac's equation and wanted to study how \
Dirac's

formalism might be affected by gravity. The rationale for doing so is \
given in

Schr\"{o}dinger's paper and will be discussed further elsewhere. In this
paper we only

discuss Schr\"{o}dinger's results in the light of their relevance to the
Madelung rule and its

anomalies. To squeeze our presentation to a minimum, we follow

(Kay, 2020). We begin with the Dirac equation 
\begin{equation}
i\gamma ^{a}\partial _{a}\psi -m\psi =0  \tag{26a}
\end{equation}

in which Dirac gamma matrices $\gamma ^{a}$ obey the Clifford algebra
anticommutation rule :

$\gamma ^{a}\gamma ^{b}+\gamma ^{b}\gamma ^{a}=2\eta ^{ab},$ $a=1\div 4,\eta
^{ab}$ is the matrix enforcing the Minkowski signature

\{1,-1,-1,-1\}. As is well known, the equivalence principle of general
relativity locally

allows us to eliminate the effects of gravity (e.g. recall the falling
elevator gedanken

experiment). Mathematically, this can be achieved by introduction of a
vierbein $e_{\mu }^{a}(x)$

so that $e_{\mu }^{a}(x)e_{\nu }^{b}(x)\eta _{ab}=g_{\mu \nu }(x)$ and $%
e_{a}^{\mu }(x)e_{b}^{\nu }(x)g_{\mu \nu }=\eta _{ab}(x).$

Thus, the vierbeins carry in themselves the effects of gravity.

To introduce these effects into Eq.(26) can be done as follows. First, the
anticommutator

$\gamma ^{a}\gamma ^{b}+\gamma ^{b}\gamma ^{a}=2\eta ^{ab}$ \ is replaced by 
$\gamma ^{\mu }\gamma ^{\nu }+\gamma ^{\nu }\gamma ^{\mu }=2g^{\mu \nu }$
implying the relationship $\gamma ^{\mu }=e_{a}^{\mu }\gamma ^{a}.$

The partial derivative $\partial _{\mu }$ is replaced \ now by the covariant
derivative%
\begin{equation}
\nabla _{\mu }\psi =\partial _{\mu }\psi +\Gamma _{\mu }\psi ,  \tag{27}
\end{equation}

where 
\begin{equation}
\Gamma _{\mu }(x)=-\frac{i}{4}\omega _{ab\mu }(x)\sigma ^{ab};\sigma ^{ab}=%
\frac{i}{2}[\gamma ^{a},\gamma ^{b}]  \tag{28}
\end{equation}

and 
\begin{equation}
\omega _{b\mu }^{a}=e_{\nu }^{a}\partial _{\mu }e_{b}^{\nu }+e_{\nu
}^{a}e_{b}^{\rho }\Gamma _{\rho \mu }^{\nu }.  \tag{29}
\end{equation}

In the simplest case $\Gamma _{\rho \mu }^{\nu }$\ is the standard
Levi-Civita connection determined by the metric

tensor $g_{\mu \nu }.$The presence of term $e_{\nu }^{a}\partial _{\mu
}e_{b}^{\nu }$ in Eq.(29) is responsible for the torsion effects.

These are absent in the canonical general relativity. Extension of general
relativity

accounting for the torsion effects is known as Einstein-Cartan\ (ECG) theory
of

gravity (Hehl et al, 1976). Very accessible (physically illuminating)
description of the

ECG theory is given in the book (Sabbata and Sivaram,1994), see also \
(Ruggiero

and Tartaglia, 2003). Above, in b), we were concerned with limitations of
special

relativity. In ECG \ theory these limitations are removed. Even more so \ in
teleparallel

gravity obtainable from ECG theory. Specifically, the Poincare group of
special

relativity is a semidirect product of Lorentz rotations and translations in
4-space\footnote{%
E.g. read\ the Poincare Group in wikipedia}.

The mass is connected with translational part of the Poincar$e^{\prime }$

group while the spin- with its rotational part. In view of this, in ECG it
is convenient to

associate with the pseudo-Riemannian space \ V$_{4}$ of the canonical
Einsteinian gravity

the space U$_{4}$ made of V$_{4}$ and its tangent space associated with
every point of V$_{4}\footnote{%
This is very much the same as considering an orthogonal frame of \ tangent
Serret-Frenet vectors moving along a curve. The vectors are moving in
Euclidean space associated with orthogonal frame while the curve is having
its local curvature and torsion and could live in non Euclidean space}.$

These arguments explain use of Latin (e.g. a,b,..)-for tangent space, and

Greek (e.g. $\mu ,\nu ,...)$ indices-forV$_{4},$ in Eq.s (27-29).Consider
now a closed curve in V$_{4}$.

If one translates tangent vector along a closed curve, after eventual
arriving

back at the starting point, the vector is not going to point into the same
direction

as it was pointing initially. If this happens, this is manifestation of
curvature of V$_{4}.$

If one is watching what is happening with the same vector in U$_{4}$ one
finds

that the curve in U$_{4}$ is not going to be closed. Furthermore the final
direction of

the tangent vector is going to be the same as that in V$_{4}$. The non
closure is associated

with torsion. Thus, the effects of curvature can be associated \ with the
effects of

Lorentz transformation of special relativity (the equivalence principle
causes this to

happen) while the effects of \ translation (associated with non
closure)-with torsion.

If in canonical gravity \ (Misner et al, 1973), chapter 7, the Einsteinian
special relativity

happens to be incompatible with \ canonical general relativity, in ECG
theory \ there is

no need to worry about a variety of frames. All frames could be equally
used. With

such a background we are ready to discuss further Schrodinger's paper of
1932. The key

element which we supply without proof is the identity%
\begin{equation}
\left( \nabla _{\alpha }\nabla _{\beta }-\nabla _{\beta }\nabla _{\alpha
}\right) \psi =\frac{1}{8}R_{\alpha \beta \delta \eta }\gamma ^{\delta
}\gamma ^{\eta }\psi .  \tag{30}
\end{equation}

Here $R_{\alpha \beta \delta \eta }$ is the Riemannian curvature tensor.
Since use of Eq.(27) converts flat space

Dirac Eq.(26a) into that in curved space%
\begin{equation}
i\gamma ^{a}\nabla _{a}\psi -m\psi =0,  \tag{26b}
\end{equation}

we can consider instead of Eq.(26b) the following equation 
\begin{eqnarray}
0 &=&(-i\gamma ^{\mu }\nabla _{\mu }\psi -m\psi )(i\gamma ^{\nu }\nabla
_{\nu }\psi -m\psi )  \notag \\
&=&\gamma ^{\mu }\gamma ^{\nu }(\nabla _{\mu }\nabla _{\nu }+\nabla _{\nu
}\nabla _{\mu }+\nabla _{\mu }\nabla _{\nu }-\nabla _{\nu }\nabla _{\mu
}+m^{2})\psi  \notag \\
&=&(g^{\mu \nu }\nabla _{\mu }\nabla _{\nu }+m^{2}+\frac{1}{8}R_{\alpha
\beta \delta \eta }\gamma ^{\mu }\gamma ^{\nu }\gamma ^{\delta }\gamma
^{\eta })\psi  \TCItag{31a}
\end{eqnarray}

obtained with use of $\gamma ^{\mu }\gamma ^{\nu }+\gamma ^{\nu }\gamma
^{\mu }=2g^{\mu \nu }$ and Eq.(30). The above equation can be further

rearranged (Moore, 1996) yielding the equivalent final result%
\begin{equation}
\left( g^{\mu \nu }\nabla _{\mu }\nabla _{\nu }+m^{2}+\frac{R}{4}\right)
\psi =0.  \tag{31b}
\end{equation}

As it was demonstrated in (Kholodenko and Kauffman, 2018) the mass term is
not

essential and can be eliminated by the appropriate substitutions. In
(Kauffman and

Kholodenko, 2019), sections 3 and 5, we \ demonstrated \ that Eq.(31b) (with 
$m=0$)

is exactly equivalent to Eq.(4). Moreover by choosing the \ potential,
Eq.(5),

(with $\gamma =1/2)$ in Eq.(4) provides the solution to L\"{o}wdin's
challenge problem\medskip , that

is establishes the ab initio validity of the canonical Madelung rule. The
obtained result,

Eq.(31b), is \ incomplete though. To make it complete, following (Schr\"{o}%
dinger, 1932)

we have to modify definition of the covariant derivative in Eq.(27) \ that
is to replace

$\nabla _{\mu }=\partial _{\mu }+\Gamma _{\mu }$ by $\nabla _{\mu }=\partial
_{\mu }+\Gamma _{\mu }-ieA_{\mu }$ where $A_{\mu }$ is some kind of a vector

(e.g. electromagnetic) potential. With such a replacement Eq.(31b) must be
substituted

by 
\begin{equation}
\left( g^{\mu \nu }\nabla _{\mu }\nabla _{\nu }+m^{2}+\frac{R}{4}+\frac{ie}{2%
}\sigma ^{ab}F_{ab}\right) \psi =0;F_{ab}=\partial _{a}A_{b}-\partial
_{b}A_{a}.  \tag{32}
\end{equation}

This is the final result obtained by Schr\"{o}dinger (up to signs and
factors i and e). These

factors can be correctly restored, e.g. \ by consulting (Berestetskii et al,
1982) page120,

Eq.(32.6). Previously obtained Eq.s(20) and (21) can be related to Eq.s(31b)
and (32).

Specifically, by putting the fine structure constant $\alpha $ in the
Lippmann-Johnson operator

to zero we are effectively arriving at Eq.(31b). For nonzero $\alpha $ we
have to use Eq.(32).\bigskip

{\Large Conclusions\bigskip }

In this work we demonstrated that ab initio solution of the Madelung rule
problem

cannot be made just with help of group-theoretic methods. The existing
Madelung

rule and its exceptions can be detected with help of the well developed to
date

Hartree-Fock calculations. However, the Madelung rule and its exceptions
carry

much more information than required for its uses in chemistry. Recall that
the

invention of quantum mechanics in 1925-1926 was driven by the needs of atomic

physics initially. \ Subsequently, quantum mechanics was extended to quantum

field theory resulting in design of the Standard Model of particle physics.
It

is exact analog of the periodic system of chemical elements\footnote{%
https://en.wikipedia.org/wiki/Standard\_Model}. Not

surprisingly, methods of particle physics had been recently applied back to

periodic system of elements (Fet, 2016),(Varlamov, 2018).

Einstein was not happy with the formalism of quantum mechanics, mainly
because

this formalism did not have room for his general relativity formalism. In
this work

we demonstrated, that the Madelung rule and its exceptions could be used for
further

development of quantum mechanical formalism because this rule and its
exceptions

are explicable by a delicate blending of the formalism of quantum mechanics
and

theory of gravitation, especially in its latest form of teleparallel
gravity. We hope,

that our work may stimulate further research in the atomic, nuclear, particle

and gravitational physics.

\bigskip

\bigskip {\Large References\bigskip \bigskip }

Allen, l., Knight, E. (2002). The L\"{o}wdin challenge. Origin of the $n+l,n$
(Madelung)\ rule

\ \ \ \ \ \ \ \ \ \ \ \ \ \ for filling the orbital configurations. \textit{%
Int.J.Quantum Chemistry} 

\ \ \ \ \ \ \ \ \ \ \ \ \ \ 90\textbf{,} 80-88.doi:10.1002/qua.965

Al-Hashimi, M.,Wiese,U. (2008) Runge-Lentz vector, accidental SU(2) symmetry,

\ \ \ \ \ \ \ \ \ \ \ \ \ \ and unusual multiplets for motion on the cone. 
\textit{Ann.Phys.}323, 82-104.

Aldrovandi, R., Pereira, J.(2013). \textit{Teleparallel Gravity}

\ \ \ \ \ \ \ \ \ \ \ \ \ \ (Springer Science+Business Media Inc.).1-214.

Aisenman, M. (1985). The intersection of Brownian paths as a case study of a

\ \ \ \ \ \ \ \ \ \ \ \ \ \ renormalization group method for quantum field
theory.

\ \ \ \ \ \ \ \ \ \ \ \ \ \ \textit{Comm.Math.Physics} 97, 91-110.

Akila, M., Waltner, D., Gutkin, B., Braun,\ P., Guhr,\ Th.(2017).\
Semiclassical

\ \ \ \ \ \ \ \ \ \ \ \ \ \ identification of periodic orbits in a quantum
many-body system.

\ \ \ \ \ \ \ \ \ \ \ \ \ \ \textit{Phys. Rev. Lett.} 118\textbf{,} 16410.

Ballesteros, A., Encio, A., Herranz, F. (2009). Hamiltonian systems
admitting a

\ \ \ \ \ \ \ \ \ \ \ \ \ \ Runge--Lenz vector and an optimal extension of
Bertrand's theorem to 

\ \ \ \ \ \ \ \ \ \ \ \ \ \ curved manifolds. \textit{Comm.Math.Phys.} 290%
\textbf{,} 1033-1049.

\ \ \ \ \ \ \ \ \ \ \ \ \ \ doi: 10.1007/s00220-009-0793-5.

Baratini, L, Christillin, P. (2012). The Machian origin of the centrifugal
force.

\ \ \ \ \ \textit{\ \ \ \ \ \ \ \ \ J.Mod.Phys}. 3, 1298-1300. doi:
10.4236/jmp.2012.329167.

Bethe, H., Jackiw,R. (2018) \textit{Intermediate Quantum Mechanics.}

\ \ \ \ \ \ \ \ \ \ \ \ \ \ (CRC Press).1-217.

Berestetskii,V., Lifshitz, E.,Pitaevski,L. (1982) \textit{Quantum
Electrodynamics}.

\ \ \ \ \ \ \ \ \ \ \ \ \ \ (Pergamon Press,Ltd.)

Biedenharn, L., Louck, J.(1981) \textit{Angular Momentum in Quantum Physics}.

\ \ \ \ \ \ \ \ \ \ \ \ \ \ (Addison-Wesley Publ.Co).1-716.

Biedenharn, L.(1983) The "Sommerfeld puzzle."Revisited and resolved.

\ \ \ \ \ \ \ \ \ \ \ \ \ \ \textit{Found.Phys}. 13, 13-34.

Born, M. (1936) \textit{The Restless Universe}. (Harper \& Brothers
Publishers).1-278.

Brihaye, Y., Kosinski, P.,Maslanka P. (2014) \ Dynamics on the cone: Closed

\ \ \ \ \ \ \ \ \ \ \ \ \ \ \ orbits and superintegrability. Ann.Phys. 344,
253-262.

Burkhardt,Ch., Leventhal, J. (2006). \textit{Topics in Atomic Physics. }

\ \ \ \ \ \ \ \ \ \ \ \ \ \ \ (Springer Science+Business Media Inc.).1-287.

Burkhardt, Ch., Hezel,T., Ciocca M., L-W.He, Leventhal,J. (1992). Classical
view of

\ \ \ \ \ \ \ \ \ \ \ \ \ \ \ \ \ the properties of Rydberg atoms:
Application of the correspondence principle.

\ \ \ \ \ \ \ \ \ \ \ \ \ \ \ \ \textit{Am.J.Physics} 60,
329-335.doi:10.1119/1.16876.

Caratheodory, C. (1937) \textit{Geometrische Optics}. (Springer, Berlin).
1-104.

Collas, P. (1978\textbf{). }Algebraic solution of the Kepler problem using
the Runge-Lentz

\ \ \ \ \ \ \ \ \ \ \ \ \ \ \ \ vector. \textit{Am.J. of Phys.}38, 253-255.

Darwin, C. (1931). \textit{The New Concepts of Matter}. (The McMillan Co.New
York).1-224.

Dirac, P. (1958). \textit{Principles of Quantum Mechanics.} (Clarendon
Press, Oxford).1-315.

Demkov,Y., Ostrovsky,V. (1971) Internal symmetry of the Maxwell "fish-eye"

\ \ \ \ \ \ \ \ \ \ \ \ \ \ \ \ \ problem and the Fock group for hydrogen
atom. \textit{\ }

\ \ \ \ \ \ \ \ \ \ \ \ \ \ \ \ \textit{Sov.Phys. JETP} 13\textbf{, }%
1083-1087.

Dyall, K., Faegri, K. (2007) \textit{Introduction to Realativistic Quantum
Chemistry}.

\ \ \ \ \ \ \ \ \ \ \ \ \ \ \ \ (Oxford U.Press). 1-518.

Englefield, M. (1972). \textit{Group Theory and Coulomb Problem. }(Wiley
Interscience).1-120.

Essen, H. (2013). The exact darwin Lagrangian. European Physics Lett. 79,
6002.

\ \ \ \ \ \ \ \ \ \ \ \ \ \ \ \ \ doi: 10.1209/0295-5075.

Fet, A. (2016). \textit{Group Theory of Chemical Elements} (de Gryuter,
Berlin).1-184.

Flugge, S. (1999). \textit{Practical Quantum Mechanics}. (Springer-Verlag,
Berlin)

Goldstein, H., Poole,C., Safko, J. (2014). \textit{Classical Mechanics.}

\ \ \ \ \ \ \ \ \ \ \ \ \ \ \ \ (Pearson Education Limited).1-630.\textit{\ }

Gorgoulhon, E. (2013). \textit{Special Relativity in General Frames}.

\ \ \ \ \ \ \ \ \ \ \ \ \ \ \ \ (Springer-Verlag,Berlin). 1-770.

Granovskii, Ya. (2004). Sommerfeld formula and Dirac's theory. 

\ \ \ \ \ \ \ \ \ \ \ \ \ \ \ \ \textit{Physics Uspekhi} 47,523-524.

Gutzviller, M. (1990).\textit{\ Chaos in Classical and Quantum Mechanics}.

\ \ \ \ \ \ \ \ \ \ \ \ \ \ \ \ (Springer-Verlag, Berlin).1-432.

Hehl, F., Heyede,P., Kerlick, G. (1976). General relativity with spin and
torsion.

\ \ \ \ \ \ \ \ \ \ \ \ \ \ \ \ \ \textit{Rev.Mod.Phys}. 48, 393-416.

Kay, B.(2020). Editorial note to: Erwin Schrodinger, Dirac electron in the
gravitational

\ \ \ \ \ \ \ \ \ \ \  \ \ \ \ \ \ field I \textit{Gen. Relativity and
Gravitation} 52(3) ,1-14.

\ \ \ \ \ \ \ \ \ \ \ \ \ \ \ \ \  doi: 10.1007/s10714-019-2625-z.

Karwowski, J., Martin, I (1991). Quantum defect orbitals and the Dirac
second-order

\ \ \ \ \ \ \ \ \ \  \ \ \ \ \ \ \ \ equation. \textit{J.Phys.B}. 24,
1539-1542.

Kennedy, F. (1972). Approximately Relativistic Interactions. \textit{%
Am.J.Phys}.40(1) 63-74.

\ \ \ \ \ \ \ \ \ \ \ \ \ \ \ \ \ \ \ doi:10.1119/1.1986448 

Keppeler, S.(2003). Semiclassical quantization rules for the Dirac and Pauli
equations.

\ \ \ \ \ \ \ \ \ \ \ \ \ \ \ \ \ \ \ \ \textit{Ann. Phys. }304, 40-71.

Kitagawara, Y., Barut, A. (1983) Period doubling in the $n+l$ filling rule
and dynamical

\ \ \ \ \ \ \ \ \ \ \ \ \ \ \ \ \ \ \ \ \ symmetry of the Demkov-Ostrovsky
atomic model.

\ \ \ \ \ \ \ \ \ \ \textit{\ \ \ \ \ \ \ \ \ \ J.Phys.B} 16\textbf{,}
3305-3322.

Kitagawara, Y., Barut, A. (1984). On the dynamical symmetry of the periodic
table:

\ \ \ \ \ \ \ \ \ \ \ \ \ \ \ \ \ \ \ \ \ II. Modified Demkov-Ostrovsky
atomic model.

\ \ \ \ \ \ \ \ \ \  \ \ \ \ \ \ \ \ \ \ \textit{J.Phys B} 17\textbf{,}
4251-4259.

Kirzhnitz, D., Lozovik, Y., Shpatkovskaya, G.(1976). Statistical model of
matter.

\ \ \ \ \ \ \ \ \ \ \ \ \ \ \ \ \ \ \ \ \ \ \textit{Sov.Phys. Uspekhi.} 18%
\textbf{, }649-672.

Kholodenko, A. (2000). Use of quadratic differentials for description of
defects and

\ \ \ \ \ \ \ \ \ \ \  \ \ \ \ \ \ \ \ \ \ textures in liquid crystals and
2+1 gravity. \textit{J.Geom.Phys}. 33, 59-102.

Kholodenko, A. (2013). \textit{Applications of Contact Geometry and Topology
in Physics.}

\ \ \ \ \ \ \ \ \ \ \ \ \ \ \ \ \ \ \ \ \ \ (World Scientific,
Singapore).1-475.

Kholodenko, A. (2017). Probabilistic vs optical interpretation of quantum
mechanics.

\ \ \ \ \ \ \ \ \ \ \ \ \ \ \ \ \ \ \ \ \ \ arXiv:1703.04674.

Kholodenko, A.,Kauffman L. (2018). Huygens triviality of the
time-independent 

\ \ \ \ \ \ \ \ \ \ \ \ \ \ \ \ \ \ \ \ \ \ Schr\"{o}dinger equation.
Applications to atomic and high energy physics.

\ \ \ \ \ \ \ \ \ \ \ \ \ \ \ \ \ \ \ \ \ \ \textit{Ann.Phys.} 390, 1-59.

Kholodenko, A., Kauffman L. (2019). How the modified Bertrand theorem
explains

\ \ \ \ \ \ \ \ \ \ \ \ \ \  \ \ \ \ \ \ \ \ regularities of the periodic
table I. From conformal invariance to Hopf

\ \ \ \ \ \ \ \ \ \ \ \ \ \  \ \ \ \ \ \ \ \ mapping. arXiv:1906.05278.

\ Kuru, \ S., Negro, J., Ragnisco, O. (2017). The Perlick system type I:
From the algebra

\ \ \ \ \ \ \ \ \ \ \ \ \ \ \ \ \ \ \ \ \ \ \ \ of symmetries to the
geometry of the trajectories.

\ \ \ \ \ \ \ \ \ \ \ \ \ \ \ \ \ \ \ \ \ \ \ \ \textit{Phys.Lett. A} 381%
\textbf{,} 3355-3363.

Landau, L., Lifshitz, E. (1960). \textit{Mechanics}. (Pergamon Press,
Oxford).1-170. 

Latter, R. (1955). Atomic energy levels for the Thomas-Fermi and
Thomas-Fermi-Dirac

\ \ \ \ \ \ \ \ \ \ \ \ \ \ \ \ \ \ \ \ \ \ \ \ \ potential. \textit{%
Phys.Rev. 9}9\textbf{,} 510-519.

Little, J.(1970). Nondegenerate homotopies of curves on the unit 2-sphere.

\ \ \ \ \ \ \ \ \ \ \ \ \ \ \ \ \ \ \ \ \ \ \ \ \textit{J.Diff.Geom}.\textbf{%
\ }4\textbf{, }339-348. 

Logunov, A.(1989).\textit{\ The Relativistic Theory of Gravitation.}

\ \ \ \ \ \ \ \ \ \ \ \ \ \ \ \ \ \ \ \ \ \ \ \ (Mir Publishers,
Moscow).1-320.

L\"{o}wdin, P-O. (1969). Some comments on periodic system of elements.

\ \ \ \ \ \ \ \ \ \ \ \ \ \ \ \ \ \ \ \ \ \ \ \ \ \ \textit{Int.
J.Quant.Chemistry} 3s, 331-334.

Luneburg, R. (1966). \textit{Mathematical Theory of Optics.}

\ \ \ \ \ \ \ \ \ \ \ \ \ \ \ \ \ \ \ \ \ \ \ \ \ (U.of California Press,
Los Angeles). 1-450.

Lusanna, L. (2019). \textit{Non Inertial Frames and Dirac Observables in
Relativity.}

\ \ \ \ \ \ \ \ \ \ \ \ \ \ \ \ \ \ \ \ \ \ \ \ \ (Cambridge U.Press,
Cambridge).1-325.

March, N. (1975). \textit{Self-Consistent Fields in Atoms}.(Pergamon Press
Ltd.).1-235.

Martin, I. (1997). The relativistic quantum defect orbital methods and some
of its

\ \ \ \ \ \ \ \ \ \ \ \ \ \ \ \ \ \  \ \ \ \ \ \ \ applications. \textit{%
Molecular Engineering} 7, 51-65.

Martin,P., Glauber, R. (1958). Relativistic theory of radiative orbital
electron capture.

\ \ \ \ \ \ \ \ \ \ \ \ \ \ \ \ \ \ \ \ \ \ \ \ \ \ \ \textit{Physical Review%
} 109, 1307-1325.

Misner,C., Wheeler, J. $\left( 1957\right) .$ Classical physics as geometry. 
\textit{Ann.Phys}.2, 525-603.

\ \ \ \ \ \ \ \ \ \ \ \ \ \ \ \ \ \ \ \ \ \ \ \ \ \ \
doi:10.1016/0003-4916(57)90049-0.

Misner, C., Torn, K., Wheeler, J. (1973). \textit{Gravitation} (Princeton
U.Press, Princeton).

\ \ \ \ \ \ \ \ \ \ \ \ \ \ \ \ \ \ \ \ \ \ \ \ \ \ \ \ \ 1-1275.

Moore, J. (1996). \textit{Lectures on Seiberg-Witten Theory.}
(Springer-Verlag, Berlin).1-105.

Ostrovsky,V.(1981). Dynamic symmetry of atomic potential.

\ \ \ \ \ \ \ \ \ \ \ \ \ \ \ \ \ \ \ \ \ \ \ \ \ \ \ \ \ \ \textit{J.Phys. B%
} 14\textbf{,} 4425-4439.

Perlick,V. (1992). Bertrand Spacetimes. \textit{Class.Quantum Grav.}
9,1009-1021.

Powers, R. (1971). Frequencies of radial oscillation and revolution as
affected by features

\ \ \ \ \ \ \ \ \ \ \ \ \ \ \ \ \ \ \ \ \ \ \ \ \ \ \ \ \ of a central
potential. 

\ \ \ \ \ \ \ \ \ \ \ \ \ \ \ \ \ \ \ \ \ \ \ \ \ \ \ \ \textit{Atti del
Convegno Mendeleeviano}, pp 235-242. 

Provost, J-P., Bracco, C. (2018). Lorentz's 1895 transformations, Einstein's

\ \ \ \ \ \ \ \ \ \ \ \ \ \ \ \ \ \ \ \ \ \ \ \ \ \ \ \ equivalence
principle and the perihelion shift of Mercury.

\ \ \ \ \ \ \ \ \ \ \ \ \ \ \ \ \ \ \ \ \ \ \ \ \ \ \ \ \textit{Eur.J.Phys.}%
39, 065602. doi: 10.1088/1361-6404/aad75d.

Rainich, G. (1925). Electrodynamics in general relativity theory. 

\ \ \ \ \ \ \ \ \ \ \ \ \ \ \ \ \ \ \ \ \ \ \ \ \ \ \ \textit{AMS
Transactions} 27, 106-136.

Rizzi, G,  Ruggiero, L. (2004). \textit{Relativity in Rotating Frames}.

\ \ \ \ \ \ \ \ \ \ \ \ \ \ \ \ \ \ \ \ \ \ \ \ \ \ \ (Kluwer Academic
Publishers, Boston).1-452.

Roseveare, N. (1982).\textit{\ Mercury's Perihelion. From Le Verrier to
Einstein.}

\ \ \ \ \ \ \ \ \ \ \ \ \ \ \ \ \ \ \ \ \ \ \ \ \ \ \ (Clarendon Press,
Oxford). 1-208.

Ruggiero, M., Tartaglia, A. (2003). Einstein--Cartan theory as a theory of
defects in 

\ \ \ \ \ \ \ \ \ \ \ \ \ \ \ \ \ \ \ \ \ \ \ \ \ \ \ space--time. \textit{%
Am.J.Phys.}71, 1303-1312. doi:10.1119/1.1596176

SabbataV., Sivaram,C. (1994). \textit{Spin and Torsion in Gravitation}.

\ \ \ \ \ \ \ \ \ \ \ \ \ \ \ \ \ \ \ \ \ \ \ \ \ \ \ (World Scientific,
Singapore).1-155.

Scerri, E., Restrepo,G. (2018) \textit{Mendeleev to Oganesson: A
Multidisciplinary Perspective}

\ \ \ \ \ \ \ \ \ \ \ \ \ \ \ \ \ \ \ \ \ \ \ \ \ \ \ \textit{\ on the
Periodic Table. }(Oxford University Press).1-315.

Schrodinger, E.(1932). The Dirac electron in a gravitational field I.

\ \ \ \ \ \ \ \ \ \ \ \ \ \ \ \ \ \ \ \ \ \ \ \ \ \ \ \ \textit{%
Sitzunsber.Preuss.Acad.Wiss. Phys.Math. K1} \ 105.

Singer,S.(2005)$.$\textit{Linearity, Symmetry and Prediction in the Hydrogen
Atom.}

\ \ \ \ \ \ \ \ \ \ \ \ \ \ \ \ \ \ \ \ \ \ \ \ \ \ \ (Springer-Verlag,
Berlin).1-275.

Sommerfeld, A.(1934). \textit{Atomic Structure and Spectral Lines}.

\ \ \ \ \ \ \ \ \ \ \ \ \ \ \ \ \ \ \ \ \ \ \ \ \ \ \ (Methuen \& Co Ltd.
London).1-375.

Steen, P.,Chang, Ch., Bostwick, J. (2019). Droplet motions fill a periodic
table.

\ \ \ \ \ \ \ \ \ \ \ \ \ \ \ \ \ \ \ \ \ \ \ \ \ \ \ \ \textit{PNAS} \ 116,
4849-4854. doi:101073/pnas.1817065116.

Symanzik, K. (1966). Euclidean quantum field theory. I. Equations for a
scalar model.

\ \ \ \ \ \ \ \ \ \ \ \ \ \ \ \ \ \ \ \ \ \ \ \ \ \ \textit{\ \ J. Math. Phys%
}. 7, 510--525.

Tietz, T. (1956). Atomic energy levels for the approximate Thomas-Fermi
Potential.

\ \ \ \ \ \ \ \ \ \ \ \ \ \ \ \ \ \ \ \ \ \ \ \ \ \ \ \ \textit{J.Chem.Phys.}
25\textbf{,} 789-790.

Tietz, T. (1968).  Contribution to the Thomas--Fermi Theory.

\ \ \ \ \ \ \ \ \ \ \ \ \ \ \ \ \ \ \ \ \ \ \ \ \ \ \ \ \textit{J.Chem.Phys}%
. 49\textbf{,} 4391-4393.

Thyssen, P. and Ceulemans, A. (2017). \textit{Shattered Symmetry}. 

\ \ \ \ \ \ \ \ \ \ \ \ \ \ \ \ \ \ \ \ \ \ \ \ \ \ (Oxford University
Press).1-498.

Varlamov,V.$\left( 2018\right) .$Group-theoretical description of periodic
system

\ \ \ \ \ \ \ \ \ \ \ \ \ \ \ \ \ \ \ \ \ \ \ \ \ \ of elements. \textit{%
Mathematical Structures and Modelling} 2,5-23.

\ \ \ \ \ \ \ \ \ \ \ \ \ \ \ \ \ \ \ \ \ \ \ \ \ \ doi:
10.25513/2222-8772.2018.2.5-23.

Wiswesser,W.(1945). The periodic system and atomic structure.

\ \ \ \ \ \ \ \ \ \ \ \ \ \ \ \ \ \ \ \ \ \ \ \ \ \ \textit{Journal of
Chemical Education} 22, 314-322.

\ \ \ \ \ \ \ \ \ \ \ \ \ \ \ \ \ \ \ \ \ \ \ \ \ \ doi:10.1021/ed022p314.

\ Wheeler, J.(1971). From Mendeleev's atom to the collapsing star. 

\ \ \ \ \ \ \ \ \ \ \ \ \ \ \ \ \ \ \ \ \ \ \ \ \ \ \textit{Atti del
Convegno\ Mendeleeviano}, pp 189-233.

\ Wheeler, J.(1976). Semi-classical analysis illuminates the connection \
between

\ \ \ \ \ \ \ \ \ \ \ \ \ \ \ \ \ \ \ \ \ \ \ \ \ \ potential and bound
states and scattering. \textit{Studies in}

\ \ \ \ \ \ \ \ \ \ \ \ \ \ \ \ \ \ \ \ \ \textit{\ \ \ \ Mathematical
Physics(Princeton U.Press).}\ pp. 351-422. 

\ \ \ \ \ \ \ \ \ \ \ \ \ \ \ \ \ \ \ \ \ \ 

\bigskip

\bigskip

\bigskip

\end{document}